\documentstyle[graphicx,12pt]{article}

\newcommand{\sech}{\mbox{sech}}
\newcommand{\CL}{{\cal L}}
\newcommand{\CO}{{\cal O}}
\newcommand{\CE}{{\cal E}}
\newcommand{\CH}{{\cal H}}
\begin{document}

\hspace{\fill}KL--TH 97/3
{\LARGE\bf
\begin{center}
Skyrme Sphalerons of an $O(3)$--$\sigma$ Model and 
the Calculation
of Transition Rates at Finite Temperature
\end{center}}
\vspace{9mm}
\begin{center}
{\large F. Zimmerschied\raisebox{0.8ex}{\small a,}\footnote{E--mail:
zimmers@physik.uni--kl.de},
D.H. Tchrakian\raisebox{0.8ex}{\small b,c,}\footnote{E--mail:
tchrakian@ailm.may.ie} and 
H.J.W. M\"uller--Kirsten\raisebox{0.8ex}{\small a,}\footnote{E--mail:
mueller1@gypsy.physik.uni--kl.de}} \\[1cm] {\it
\raisebox{0.8ex}{\small a}
Department of Physics \\ University of
Kaiserslautern, P.\ O.\ Box 3049, D 67653 Kaiserslautern, Germany \\[5mm]
\raisebox{0.8ex}{\small b} Department of Mathematical Physics \\ St.\
Patrick's College, Maynooth, Ireland \\[5mm] \raisebox{0.8ex}{\small c}
School of Theoretical Physics, Dublin Institute for Advanced Studies, \\ 10
Burlington Road, Dublin 4, Ireland} \\
\vfill {\bf Abstract}
\end{center}
The reduced $O(3)$--$\sigma$ model with an $O(3)\rightarrow O(2)$ symmetry
breaking potential 
is considered with an additional Skyrmionic term, i.e.\ a
totally antisymmetric quartic term in the field derivatives.
This Skyrme term does not affect the classical static equations of motion
which, however, allow an unstable sphaleron solution. 
Quantum fluctuations around the static classical
solution are considered for the determination of the rate of thermally
induced transitions between topologically distinct vacua mediated
by the sphaleron. The main technical effect of the Skyrme term is to produce
an extra measure factor in one of the fluctuation path 
integrals which is therefore evaluated using a measure--modified 
Fourier--Matsubara decomposition (this being one of the few cases 
permitting this explicit calculation). 
The resulting transition rate is valid in a temperature region different 
from that of the original Skyrme--less model, and the crossover from 
transitions dominated by thermal fluctuations to those dominated by tunneling  
at the lower limit of this range depends on the strength of the
Skyrme coupling.
\\[5mm]
\newpage

\setcounter{footnote}{0}
\renewcommand{\thefootnote}{\fnsymbol{footnote}} 

\section{Introduction}

The complicated vacuum structure of quantum field theories involving nonabelian
gauge symmetries has been a subject of considerable interest since its 
consequences have been discussed for the first time \cite{tHooftLett,tHooft}. 
In such theories, besides the usual quantum states built perturbatively on the
topologically distinct vacua (characterized by a topological quantity, the 
winding number), there are nonperturbative phenomena relating these separate
perturbative sectors \cite{model}.
In particular, there are transitions between the states
built on neighbouring vacua which are dominated by different physical processes,
depending on the temperature of the system. These processes can be described 
semiclassically by pseudoparticles, i.e.\ classical configurations with
particle--like properties. The most important example for these effects is
the electroweak model in which winding number transitions are related to
baryon/lepton number nonconservation \cite{couple,Adler}. 
Most calculations, however, are performed in 
simpler, more manageable ``toy models''.

This article presents the evaluation of such a transition rate in a special 
temperature region for a particular but typical model which has the advantage
of permitting explicit calculation in a nontrivial case. 
In this introduction, we therefore
first present the general physical background, i.e.\ the physics of transitions
between topologically distinct sectors of a quantum field theory in different 
temperature regions, and then introduce the particular model which is motivated
by its special temperature behaviour, but also by some special technical 
properties.

\subsection{Transition rates due to tunneling or thermal fluctuations}

Depending on the temperature of the system, transition processes between
neighbouring perturbative sectors of a quantum field theory are dominated by
different physical processes:

At zero temperature $T=0$, the relevant process is {\em tunneling} through the 
energy barrier separating the neighbouring vacua. In a semiclassical 
approximation, this tunneling can be described by {\em vacuum instantons}
\cite{Belavin,Fateev}. 
Vacuum instantons are pseudoparticles which are stable classical solutions of 
the Euclidean Euler--Lagrange equations of the theory with vanishing Euclidean 
energy as boundary condition, $\CE_{inst}=0$. A semiclassical expansion of the 
Euclidean path integral around the instanton yields the rate of 
tunneling through 
the barrier separating the topologically distinct vacua between which the 
instanton interpolates as a funtion of imaginary time, used here to 
describe tunneling. The tree approximation of the instanton tunneling transition
rate is thus $\exp\{-2S_0\}$ where $S_0$ is the Euclidean action of the
vacuum instanton \cite{Gildener}.

At finite temperatures $T>0$, 
there are thermally excited states in the neighbouring
sectors, and starting from some lower temperature limit, transitions due to
{\em thermal fluctuations} over the separating barrier become relevant. These 
transitions are described semiclassically by pseudoparticles called 
{\em sphalerons}. These are classical solutions of the static 
Euler--Lagrange equations with finite Euclidean energy $\CE_{sph}$ which are
unstable, i.e.\ the static Gaussian fluctuation operator around these solutions
has one negative eigenvalue $\omega_-$ \cite{Taubes,MantonKlink}. 
Sphalerons can be visualized as
``sitting on top'' of the energy barrier separating neighbouring vacua. 
The corresponding semiclassical transition rate is suppressed by the classical
Boltzmann factor, $\exp\{-\CE_{sph}\beta_T\}$ which one obtains as the tree
approximation of the partition function path integral which again involves the 
use of complex time, now in the convenient formalism of quantum field
theory at finite temperature \cite{Kleinert,FouMats}. 
Comparing the leading exponential factors
of the semiclassical expansion around instantons and sphalerons, respectively, 
one finds that thermal fluctuations dominate vacuum tunneling in the
transition rate for temperatures $T>T_0=:\frac{1}{k_B}\frac{\CE_{sph}}{2S_0}$.

A third transition process which has been discussed intensively in 
recent literature
\cite{Habib,Kuznetsov} 
is {\em tunneling from (thermally) excited states}, described by a
Boltzmann average over {\em nonvacuum instantons} which are classical solutions of the Euclidean Euler--Lagrange equations with Euclidean energy $\CE_{inst}>0$
as boundary condition \cite{Khlebnikov,LiangMK}. 
These solutions are periodic in the imaginary time
coordinate with energy dependent period $\hat{\tau}=\hat{\tau}(\CE_{inst})$
(which is why they are usually called {\em ``periodic instantons''}). In order 
to apply the Boltzmann average over the periodic instantons, one has to satisfy
the saddle point condition $\hat{\tau}(\CE_{inst})\equiv \beta_T$, so that only
periodic instantons with period equal to the inverse temperature dominate the
nonvacuum tunneling transition rate. The Euclidean action of the periodic 
instanton can thus be written $S_{inst}(\beta_T)$, and the usual relation
$\CE_{inst}=\frac{S_{inst}(\beta_T)}{\beta_T}$ holds. The two different uses of 
complex time (in tunneling and in thermodynamics) 
thus merge together in the case of the 
periodic instanton. 

For $\CE_{inst}\rightarrow\CE_{sph}$ the imaginary time dependence vanishes
and the periodic instanton reduces to the sphaleron (being time periodic with
any period, i.e.\ static) with Euclidean action 
$S_{sph}(\beta_T)=\CE_{sph}\beta_T$. The second limit $\CE_{inst}\rightarrow 0$
which should reduce the periodic instanton to the vacuum one shows that there 
are two different types of periodic instantons \cite{Kuznetsov}: Those with
$\hat{\tau}(\CE_{inst})\stackrel{\CE_{inst}\rightarrow 0}{\longrightarrow}
\infty$ where one has {\em localised vacuum instantons} with infinite period
and those with 
$\hat{\tau}(\CE_{inst})\stackrel{\CE_{inst}\rightarrow 0}{\longrightarrow} 0$ 
where localised vacuum instantons do {\em not} exist. Nonetheless the
well--known method of {\em constrained instantons} \cite{Khlebnikov,Affleck}
(which may be interpreted as the 
zero period limit of this second type of periodic
instantons) allows vacuum tunneling even in this case. In either case the
dilute gas approximation works and one has 
$S_{inst}(\beta_T)\stackrel{\beta_T\rightarrow 0}{\longrightarrow} 2S_0$.

It is now an interesting and important question to ask 
how the temperature domains
in which the different physical processes (vacuum tunneling, thermal tunneling,
thermal fluctuations) dominate the transition rate, can be continued into each 
other.The crucial point here is the crossover from predominantly 
thermal fluctuations
to predominantly 
(vacuum or thermally excited) tunneling. Near the sphaleron energy,
$\CE_{sph}-\CE_{inst}\ll\CE_{inst}$, the periodic instanton can be written as a
sum of the sphaleron and oscillations in its eigenmode with negative eigenvalue
$\omega_-$. In the limit $\CE_{inst}\rightarrow\CE_{sph}$, the period of 
oscillation approaches $\hat{\tau}_{sph}:=\frac{2\pi}{\omega_-}=:
\frac{1}{k_BT^{(-)}}$, the periodic instanton thereby merging into the sphaleron
which has Euclidean action $S_{sph}(\hat{\tau}_{sph})=S_{sph}^{(-)}$
at this lower temperature bound. 
Depending on the model, the period $\hat{\tau}(\CE_{inst})$ for 
$\CE_{inst}<\CE_{sph}$  may be smaller or larger than $\hat{\tau}_{sph}$.

\begin{figure}[t]
\begin{center}
{\includegraphics[scale=1.0]{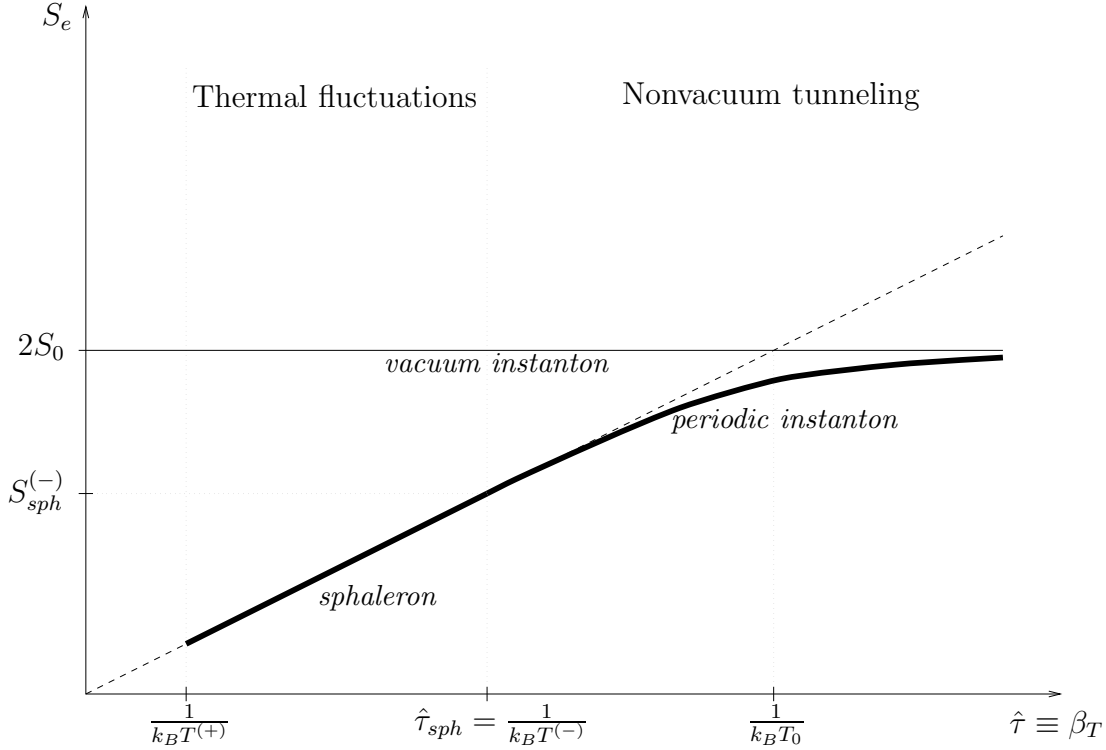}}
\put(-380,260){\makebox(0,0){$S_e$}}
\put(-2,-8){\makebox(0,0){$\hat{\tau}\equiv\beta_T$}}
\put(-388,80){\makebox(0,0){$S_{sph}^{(-)}$}}
\put(-386,135){\makebox(0,0){$2S_0$}}
\put(-330,-8){\makebox(0,0){$\frac{1}{k_BT^{(+)}}$}}
\put(-108,-8){\makebox(0,0){$\frac{1}{k_BT_0}$}}
\put(-212,-8){\makebox(0,0){$\hat{\tau}_{sph}=\frac{1}{k_BT^{(-)}}$}}
\put(-275,230){\makebox(0,0){Thermal fluctuations}}
\put(-110,230){\makebox(0,0){Nonvacuum tunneling}}
\put(-105,107){\makebox(0,0){\small\it periodic instanton}}
\put(-214,129){\makebox(0,0){\small\it vacuum instanton}}
\put(-259,40){\makebox(0,0){\small\it sphaleron}}
\end{center}
\caption{Type I theory (the solution that dominates the transition
rate is indicated by the thick line)} 
\label{figregions1}
\end{figure}
\nopagebreak

\begin{figure}[t]
\begin{center}
{\includegraphics[scale=1.0]{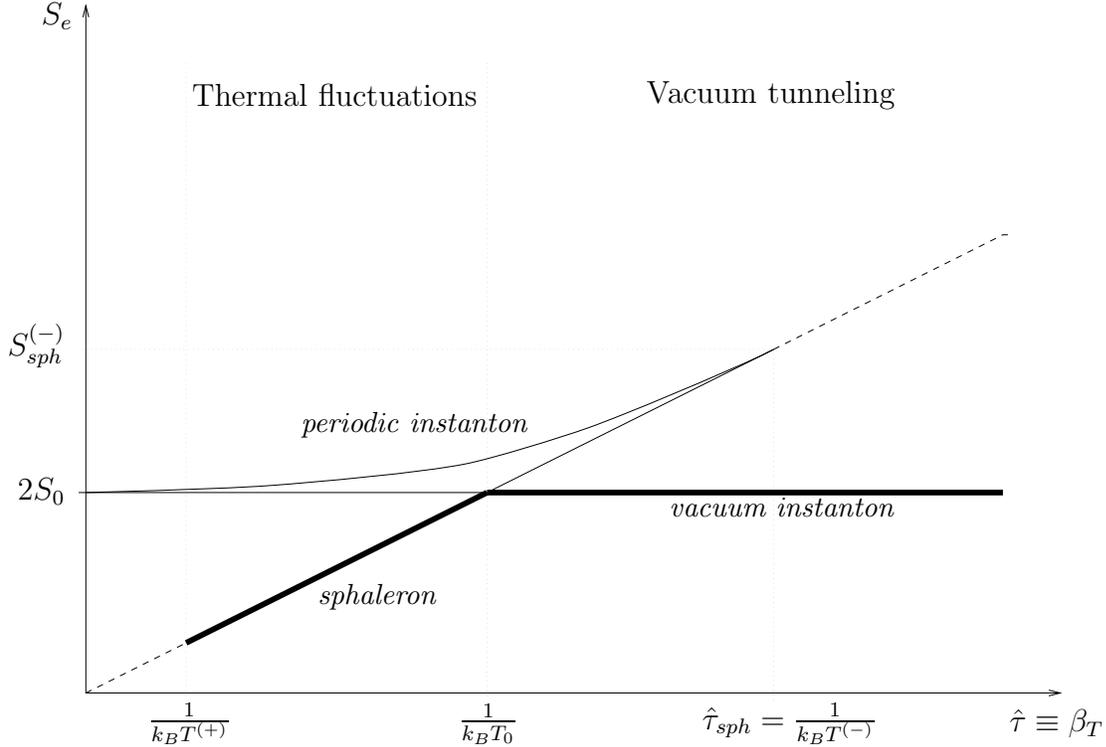}}
\put(-380,260){\makebox(0,0){$S_e$}}
\put(-2,-8){\makebox(0,0){$\hat{\tau}\equiv\beta_T$}}
\put(-388,135){\makebox(0,0){$S_{sph}^{(-)}$}}
\put(-386,80){\makebox(0,0){$2S_0$}}
\put(-330,-8){\makebox(0,0){$\frac{1}{k_BT^{(+)}}$}}
\put(-217,-8){\makebox(0,0){$\frac{1}{k_BT_0}$}}
\put(-103,-8){\makebox(0,0){$\hat{\tau}_{sph}=\frac{1}{k_BT^{(-)}}$}}
\put(-275,230){\makebox(0,0){Thermal fluctuations}}
\put(-110,230){\makebox(0,0){Vacuum tunneling}}
\put(-245,105){\makebox(0,0){\small\it periodic instanton}}
\put(-106,74){\makebox(0,0){\small\it vacuum instanton}}
\put(-259,40){\makebox(0,0){\small\it sphaleron}}
\end{center}
\caption{Type II theory (the solution that dominates the transition
rate is indicated by the thick line)}
\label{figregions2}
\end{figure}

This fact, together with the two types of low energy behaviour of the periodic
instantons, results in two different simple types of theories with continuous
crossover between the temperature domains with different dominant processes
\cite{Kuznetsov}.
In {\em type I theories}, we have localised instantons, 
$\hat{\tau}(\CE_{inst})\stackrel{\CE_{inst}\rightarrow 0}{\longrightarrow}
\infty$, the period $\hat{\tau}$ is always greater than $\hat{\tau}_{sph}$,
thus $\frac{\partial^2 S_{inst}(\beta_T)}{\partial \beta_T^2}=
\frac{\partial\CE_{inst}}{\partial \beta_T}<0$, and 
$2S_0>S_{sph}^{(-)}$. Then $T_0<T^{(-)}$, thermal fluctuations
dominate for $T>T^{(-)}$ and nonvacuum tunneling from thermally excited states
dominates for $T<T^{(-)}$, and is suppressed by $\exp\{-S_{inst}(\beta_T)\}$.
The crossover between the two regions is smooth. This situation is shown in 
Fig.\ \ref{figregions1}.
{\em Type II theories} have nonlocalised vacuum instantons,  
$\hat{\tau}(\CE_{inst})\stackrel{\CE_{inst}\rightarrow 0}{\longrightarrow} 0$,
the period $\hat{\tau}$ is always smaller than $\hat{\tau}_{sph}$, thus
$\frac{\partial^2 S_{inst}(\beta_T)}{\partial \beta_T^2}=
\frac{\partial\CE_{inst}}{\partial \beta_T}>0$, and 
$2S_0<S_{sph}^{(-)}$. This yields $T^{(-)}<T_0$: 
Thermal fluctuations
dominate for $T>T_0$, vacuum tunneling for $T<T_0$ and the crossover is sharp
as shown in Fig.\ \ref{figregions2}.

Well known examples are the $1+1$ dimensional
Higgs model \cite{Khlebnikov,Bochka} 
for a type I theory and the reduced nonlinear
$O(3)$--$\sigma$ model \cite{MottWipf,Habib} for a type II theory. Both are
prominent toy models for the winding number transition in the electroweak model.
But the electroweak theory seems to be neither of type I nor of type II: It has
no localised instantons which disagrees with type I properties, and one 
estimates $2S_0>\CE_{sph}\hat{\tau}_{sph}$ for the electroweak constrained 
instantons, so that the theory is not of type II.

To understand this special property of the electroweak model which is not shared
by the usual toy models, it is interesting to study simple low dimensional 
models which also can not be classified under 
the two types above, although such 
models may not be good toy models for the electroweak theory from other points 
of view.
One of these models of recent interest \cite{Kuznetsov} is the reduced nonlinear
$O(3)$--$\sigma$ model in $1+1$ dimensions with an added Skyrme term 
\cite{modelidea}.

\subsection{The reduced $O(3)$--$\sigma$ model with Skyrme term and its
motivation}

The model we deal with here is defined by the Lagrangian
density
\begin{eqnarray}
\CL(\phi^a,\partial_{\mu}\phi^a) & = &
\frac{\lambda_2}{8}(\partial_{[\mu}\phi^a\partial_{\nu]}\phi^b)
(\partial^{[\nu}\phi^a\partial^{\mu]}\phi^b)
+\frac{\lambda_1}{2} 
(\partial_{\mu}\phi^a\partial^{\mu}\phi^a) - \lambda_0 V(\phi^3) \nonumber \\
& = & 
\frac{\lambda_2}{2}\left[\dot{\vec{\phi}}\wedge\vec{\phi}'\right]^2
+\frac{\lambda_1}{2}\left[(\dot{\vec{\phi}})^2 -
(\vec{\phi}')^2\right] - \lambda_0 V(\phi^3) \nonumber \\
& = & 
\lambda_1\left\{\frac{\nu^2}{2}\left[(\dot{\vec{\phi}})^2(\vec{\phi}')^2 -
(\dot{\vec{\phi}}\cdot\vec{\phi}')^2 
\right]+\frac{1}{2}\left[(\dot{\vec{\phi}})^2 -
(\vec{\phi}')^2\right] - \mu^2 V(\phi^3)\right\} 
\label{1}
\end{eqnarray}
with
\begin{equation}
\nu^2:=\frac{\lambda_2}{\lambda_1},\qquad \mu^2:=\frac{\lambda_0}{\lambda_1}
\label{1a}
\end{equation}
in Minkowski spacetime $(x^0,x^1)=(t,x)$ with fields 
$\vec{\phi}=(\phi^a)$,
$a=1,2,3$ constrained by $\phi^a\phi^a=1$ (we do not add a Lagrange multiplier
since we will always work with parametrisations respecting the constraint).
This Lagrangian differs from that of the usual nonlinear $\sigma$ model,
$\CL=\frac{1}{2g^2}(\partial_{\mu}\phi^a\partial^{\mu}\phi^a)$ \cite{JETP}
(comparing the kinetic terms of the two Lagrangians, the two couplings
$\frac{1}{g^2}=\lambda_1$ have to be identified) with
$\phi^a\phi^a=1$ by two additional terms. The explicit $O(3)\rightarrow O(2)$
symmetry breaking potential (which we choose to be
$V(\phi^3)=1-\phi^3$), being necessary to produce a sphaleron in the theory
(there is no mechanism of spontaneous symmetry breaking for $\sigma$ models),
and the Skyrme term with coupling $\nu^2$. 
Setting $\nu^2=0$ reduces this model to
the well--known reduced nonlinear $O(3)$--$\sigma$ model which provides good
model predictions for the electroweak theory \cite{MottWipf}. The latter is the
original Skyrme--less model which has {\em no} localised instantons 
and is a type II theory \cite{Habib}.

Adding a Skyrme term changes this last aspect completely. 
First, there is a localised 
vacuum instanton $\varphi_{inst}$ in the Skyrmed theory \cite{OBrien}
which can be 
verified numerically \cite{modelidea}. Its action is \cite{Kuznetsov}
\begin{equation}
S_0 = 4\pi\lambda_1 + \CO(\lambda_1\nu\mu).
\label{1b}
\end{equation}

In this model the sphaleron solution 
$\varphi_{sph}$ is not affected by the Skyrme term because it solves the
{\em static} Euler--Lagrange equations and the Skyrme term is purely dynamical.
But the fluctuation spectrum around the sphaleron is modified and therefore
also the one--loop quantum correction in the semiclassical transition rate.
Moreover, the temperature range of validity of this calculation changes. 
In particular,
the temperatures $T_0$ and $T^{(-)}$ will depend on $\nu$.

It is the purpose of this investigation 
to calculate the thermal transition rate in a
semiclassical one--loop expansion around the sphaleron in this Skyrmed theory. 
This involves the evaluation of fluctuation determinants, and one main 
motivation of this paper (besides knowing that the Skyrmed reduced nonlinear 
$O(3)$--$\sigma$ model is a theory neither of type I nor of 
type II)
is that these evaluations can be performed 
explicitly, and one can finally see clearly the role
of the Skyrme term in the thermal transition rate. 

Moreover, the technical aspects of the calculational difficulties which
the Skyrme term gives rise to are presented in detail. 
These techniques should be of interest also in the context of
more physically motivated models like the three dimensional 
$\sigma$ model which is motivated by QCD. In the case of the latter, 
the hope is that the 
loop contributions could give rise to Skyrme--like terms in the action which 
would stabilize the Skyrmion \cite{Adkins}.

The layout of the paper is as follows: In {\em Section \ref{section1}} 
the notion of a sphaleron is reviewed. The NCL technique is discussed
in the context of the Skyrmed $O(3)$--$\sigma$ model, the sphaleron solution 
is derived and its instability is discussed in terms of the
static Gaussian fluctuation operators. 

The embedding of the sphaleron of the reduced  $O(3)$--$\sigma$ model into the
full Euclidean theory is described in {\em Section \ref{section2}} where the 
thermodynamical formalism yielding the transition rate is reviewed, 
using the path integral version of quantum statistics \cite{Feynman}. What 
is quantized are the fluctuations parallel and perpendicular to the vacuum 
and the sphaleron, respectively. Since these fluctuations are dynamical 
quantities, they may involve the (kinetic) Skyrme term. The effect of the 
latter on the model is to produce an extra measure factor which 
modifies the scalar product in the Hilbert space of the perpendicular
fluctuations around the sphaleron. Therefore the evaluation of the path 
integrals is also considered in some detail in this Section, taking care of 
the path integral measures.

{\em Sections \ref{sectionvacfluc} and \ref{sectionparfluc}} deal with the 
fluctuations around the vacuum and the parallel fluctuations around the 
sphaleron which are already well known since there is no influence of the 
Skyrme term on this part of the theory \cite{MottWipf}.

The new effects of the Skyrme measure factor on the perpendicular fluctuations
around the sphaleron are investigated in {\em Section \ref{section5}}.

In {\em Section \ref{sectionrange}} the results of the calculations are
summarized to yield the thermally activated transition rate. The Skyrme--less
limit is discussed, as well as the temperature range of validity of the 
calculation. 

The physical implications of this analysis are then discussed in the concluding
{\em Section \ref{sectionconc}}.

Finally, some mathematical steps and calculational details are given in 
the {\em Appendices}.

\section{Saddle--point configurations of the static energy functional}

\label{section1}

The solutions we consider are static, $\vec{\phi}=\vec{\phi}(x)$, 
and result as
extrema of the static energy functional
\begin{equation}
{\cal E}[\vec{\phi}] = \int {\cal H}_{stat} dx
\label{1.1}
\end{equation}
where ${\cal H}_{stat}$ is the static part of the Hamiltonian, 
defined as the $00$--component ${\cal H}=T_{00}$ of the energy--momentum tensor
$T_{\mu\nu}:=\Pi^a_{\mu}\phi^a_{\nu}-g_{\mu\nu}\CL $, 
$\Pi^a_{\mu}:=\frac{\partial\CL }{\partial(\partial^{\mu}\phi^a)}$, or 
obtained alternatively from the well--known formula 
\begin{equation}
{\cal H}:=\vec{\Pi}\dot{\vec{\phi}}-
\CL , \qquad \vec{\Pi}:=
\frac{\partial \CL }{\partial \dot{\vec{\phi}}}
\label{1.1a}
\end{equation}
Although one cannot write down the Hamiltonian in terms of
fields and conjugate momenta explicitly, it is easy to see that
\begin{equation}
{\cal H}_{stat} = \frac{\lambda_1}{2}(\vec{\phi}')^2+\lambda_0V(\phi^3).
\label{1.2}
\end{equation}
For the special potential $V(\phi^3)=1-\phi^3$, the vacuum configuration
of $\CE[\vec{\phi}]$ is thus given by $\vec{\varphi}_0=(0,0,1)$.

We parametrize the fields so that they satisfy the constraint $\vec{\phi}^2=1$.
There are at least three appropriate parametrizations of the static fields,
\begin{eqnarray}
\vec{\phi}(x) & = & \frac{1}{\sqrt{1+(f^2(x))^2}}\left(\begin{array}{c} 
\sin f^1(x) \\ f^2(x) \\
\cos f^1(x) \end{array} \right)
\label{1.4a} \\
\vec{\phi}(x) & = & \left(\begin{array}{c} 
\sin g^1(x) \cos g^2(x) \\ \sin g^1(x) \sin g^2(x) \\
\cos g^1(x) \end{array} \right)
\label{1.4} \\
\vec{\phi}(x) & = & \left(\begin{array}{c} 
\sin h^1(x) \sin h^2(x) \\ \sin h^1(x) \cos h^1(x) (\cos h^2(x)-1)\\
\sin^2 h^1(x)\cos h^2(x) + \cos^2 h^1(x)\end{array} \right)
\label{1.3} 
\end{eqnarray}
In terms of the new parameter fields $\vec{f}(x)$, $\vec{g}(x)$ and
$\vec{h}(x)$, the static
energy functional (\ref{1.1}) is $\CE[\phi] = \CE_f[\vec{f}] = 
\CE_g[\vec{g}] = \CE_h[\vec{h}]$ with
\begin{eqnarray}
\CE_f[\vec{f}]  
& = & \int \left[ \lambda_1\langle 
\vec{f}'(x),F(\vec{f}(x))\vec{f}'(x)\rangle + \lambda_0
V\left(\frac{\cos f^1(x)}{\sqrt{1+(f^2(x))^2}}\right) \right] dx
\label{1.5} \\
\CE_g[\vec{g}] 
& = & \int \left[ \lambda_1\langle 
\vec{g}'(x),G(\vec{g}(x))\vec{g}'(x)\rangle + \lambda_0
V\left(\cos g^1(x)\right)\right] dx \label{1.6} \\
\CE_h[\vec{h}] 
& = & \int \left[ \lambda_1\langle 
\vec{h}'(x),H(\vec{h}(x))\vec{h}'(x)\rangle + \lambda_0
V\left(\sin^2 h^1(x)\cos h^2(x) + \cos^2 h^1(x)\right)\right] dx 
\nonumber \\ \label{1.6a} 
\end{eqnarray}
with the matrices
\begin{eqnarray}
F(\vec{f}) & = & \left(\begin{array}{cc} 
\frac{1}{1+(f^2)^2} & 0 \\ 0 & 
\frac{1}{(1+(f^2)^2)^2}
\end{array} \right) 
\label{1.7} \\
G(\vec{g}) & = & \left(\begin{array}{cc} 
1 & 0 \\ 0 & \sin^2 g^1 \end{array} \right) \label{1.8} \\
H(\vec{h}) & = & \left(\begin{array}{cc}
\cos^2 h^1 \sin^2 h^2 + (1-\cos h^2)^2 & \sin h^1 \cos h^1
\sin h^2 \\ \sin h^1 \cos h^1 \sin h^2 & \sin^2 h^1
\end{array} \right) \nonumber \\ \label{1.8a}
\end{eqnarray}

The $\vec{h}$--parametrization is usually used to find the saddle--point
of the static energy functional (\ref{1.1}) with the special potential
$V(\phi^3)=1-\phi^3$. Treating $h^1(x)$ as a
constant parameter $h^1(x)\equiv\eta$ and writing $h^2(x)=f(x)$, the
parametrization (\ref{1.3}) yields a noncontractible loop (NCL) \cite{NCL}
\begin{equation}
\vec{\phi}_{\eta}(x)  =  \left(\begin{array}{c} 
\sin \eta \sin f(x) \\ \sin \eta \cos \eta (\cos f(x)-1)\\
\sin^2 \eta\cos f(x) + \cos^2 \eta\end{array} \right)
\label{1.9} 
\end{equation}
in the space of {\em static} field configurations on which $\CE$ is
defined. The NCL starts and ends in the vacuum.
The energy of the configurations along this loop is given by
\begin{equation}
\CE[\vec{\phi}_{\eta}] = \lambda_1\sin^2 \eta \int \left[
\frac{1}{2}(f'(x))^2+\mu^2(1-\cos f(x))\right] dx
\label{1.10}
\end{equation}
which has a maximum in the $\eta$--direction for $\eta=\frac{\pi}{2}$.
Minimizing the resulting energy function by variation of $f(x)$,
\begin{equation}
\delta \CE[\vec{\phi}_{(\eta=\frac{\pi}{2})}]
=0 \quad
\Longrightarrow \quad \lambda_1\frac{\delta}{\delta f(y)}
\int \left[
\frac{1}{2}(f'(x))^2+\mu^2(1-\cos f(x))\right] dx = 0
\label{1.11}
\end{equation}
yields the
well--known Sine--Gordon equation
\begin{equation}
f''(x) - \mu^2\sin f(x) = 0, \qquad \mu^2 := \frac{\lambda_0}{\lambda_1}.
\label{1.12}
\end{equation} 
Assuming as boundary condition that for $|x|\rightarrow\infty$, $\vec{\phi}(x)$
approaches the vacuum $\vec{\varphi}_0$, i.\ e.\ 
$f(x)\stackrel{|x|\rightarrow\infty}{\longrightarrow}0,2\pi$, the solution of 
(\ref{1.12}) is
\begin{equation}
f(x)=4 \arctan\left(e^{\pm\mu(x-x_0)}\right).
\label{1.13}
\end{equation}
 
From the NCL technique \cite{NCL}, we expect the classical solution 
$\vec{\varphi}$ which 
in this parametrization is given by $\eta=\frac{\pi}{2},f(x)$,
i.e.\ $\vec{\varphi}_{sph}(x)=\vec{\varphi}_1(x)=(\sin f(x),0,\cos f(x))$, 
to be a saddle--point of
the energy functional $\CE[\vec{\phi}]$. This can be verified 
by considering
the second variational derivative of $\CE[\vec{\phi}]$ which should be
taken in a convenient 
parametrization of $\vec{\phi}$ to avoid separate treatment
of the $\vec{\phi}^2=1$ constraint. Using
the parametrizations (\ref{1.4a},\ref{1.4}, \ref{1.3}) 
we thus have to analyse the eigenvalues
of the second variation of $\CE_f[\vec{g}]$, $\CE_g[\vec{f}]$ or
$\CE_h[\vec{h}]$
taken at the classical solution $\vec{\varphi}_1(x)$, which means
$f^1(x)=f(x)$, $f^2(x)=0$; $g^1(x)=f(x)$, $g^2(x)=0$
or $h^1(x)=\frac{\pi}{2}$, $h^2(x)=f(x)$
respectively:
\begin{eqnarray}
\int \left[\frac{\delta^2 \CE_f[\vec{f}]}{\delta f^i(x) \delta f^j(x')}
\right]_{\vec{\varphi}}\psi_f^j(x') \, dx' & = & \omega^2\psi_f^i(x)
\label{1.15} \\
\int \left[\frac{\delta^2 \CE_g[\vec{g}]}{\delta g^i(x) \delta g^j(x')}
\right]_{\vec{\varphi}}\psi_g^j(x) \, dx' & = & \omega^2\psi_g^i(x)
\label{1.14} \\
\int \left[\frac{\delta^2 \CE_h[\vec{h}]}{\delta h^i(x) \delta h^j(x')}
\right]_{\vec{\varphi}}\psi_h^j(x) \, dx' & = & \omega^2\psi_h^i(x)
\label{1.15a}
\end{eqnarray}
with $i,j=1,2$.
These eigenvalue equations are equivalent. The $\vec{f}$--parametrization
yields directly decoupled equations in a simple form. Eqs.\ (\ref{1.15a}) have 
to be decoupled, whereas (\ref{1.14}) are decoupled, but can be simplified
by a substitution. This is why the  $\vec{f}$--parametrization is preferred in
the following, where we will modify our notations slightly to
$\vec{f}=\left(f^{(\parallel)},f^{(\perp)}\right)$ to denote the two
parameter fields. We will see in the next section 
that $f^{(\parallel)}$ describes fluctuations parallel and
$f^{(\perp)}$ those perpendicular to the
sphaleron $\vec{\varphi}_1(x)$. In this new notation, each of the
eigenvalue equations (\ref{1.15}--\ref{1.15a}) may be written as a system of
two decoupled static fluctuation equations, i.e.\
\begin{eqnarray}
\hat{\CH}_{\parallel}\psi^{(\parallel)}(x) 
& = & \frac{\left(\omega^{(\parallel)}\right)^2}{\lambda_1} 
\psi^{(\parallel)}(x), \qquad
\hat{\CH}_{\parallel} = - \frac{\partial^2}{\partial x^2} +
\mu^2 \left\{1-2\mbox{sech}^2\left(\mu x\right)\right\} 
\label{1.16} \\
\hat{\CH}_{\perp}\psi^{(\perp)}(x) & = & 
\frac{\left(\omega^{(\perp)}\right)^2}{\lambda_1} 
\psi^{(\perp)}(x), \qquad
\hat{\CH}_{\perp}  = - \frac{\partial^2}{\partial x^2} +
\mu^2 \left\{1-6\mbox{sech}^2\left(\mu x\right)\right\}.
\label{1.17}
\end{eqnarray}
The second equation (\ref{1.17}) has one negative eigenvalue 
$\left(\omega^{(\perp)}\right)^2=-3\mu^2$;
thus the static solution $\vec{\varphi}_1(x)$ 
is {\em always unstable in the space of
static field configurations}. It is important to note 
that we never left this space. The sphaleron is a
static object, and one can not decide whether it is Minkowskian or
Euclidean.

The notion of a ``static fluctuation operator'' reflects the fact that one can 
derive the stability operator  (\ref{1.5}) (which is usually called the
``static Gaussian fluctuation operator'') by expanding $\CE_f[\vec{f}]$
in small ``fluctuations'' (which are {\em static} in this context!)
around the classical solutions, and similary for the other parametrizations.
These fluctuations become {\em dynamical} quantities if one embeds the 
sphaleron into the full, time--dependent theory, either in its Minkowskian 
version (\ref{1}) or in the Euclidean version which one obtains from the former
by a Wick rotation in the time coordinate, $t\mapsto \tau=-it$.
As the sphaleron itself does not depend on time even after the embedding
into the full, e.g.\ Euclidean theory, one may impose periodic boundary
conditions in the imaginary time coordinate on the fields. Then one can treat 
the result as a field theory at finite temperature which is defined via the
imaginary time period $\beta_T=\frac{1}{k_BT}$ (we set $k_B=1$), and the usual
path integral 
formalism of quantum statistics \cite{Feynman} 
is applicable, including the calculation
of thermally activated transitions over energy barriers.

\section{Finite Temperature: Sphalerons in Euclidean spacetime}

\label{section2}

From the general theory of thermal transition rates \cite{Affleck,Transition}
, we know that transitions due to
fluctuations at finite temperature may be decribed by the imaginary part of
the free energy $F$,
\begin{equation}
\Gamma=\frac{|\varrho|}{\pi}\mbox{Im}(\beta_T F)
\label{2.1}
\end{equation}
where $\varrho$ is a damping constant. As usual, $F$ is related to the
partition function ${\cal Z}$ by ${\cal Z}=e^{-\beta_TF}$, so that
the partition function of the system at temperature
$T$ has to be computed, using the path integral version of quantum statistics
\cite{Feynman} \cite{Kleinert}, i.e.\
\begin{equation}
{\cal Z} = \int \prod_{i=1}^3 [{\rm d}{\phi}^i(\tau,x)]
\prod_{i=1}^3[{\rm d}{\Pi}^i(\tau,x)]
\delta\left(|\vec{\phi}|-1\right)
e^{-S_e[\vec{\phi},\vec{\Pi}]}.
\label{2.2}
\end{equation}
Here we start from the Hamiltonian version of the partition function path
integral \cite{Kleinert}
(we emphasize this also by the use of the symbol $[{\rm d}\ldots]$ 
for the
path measures): To obtain the Lagrangian version, one has to integrate out
the momenta in the
Euclidean action of the system,
\begin{equation}
S_e[\vec{\phi},\vec{\Pi}] = \int_0^{\beta_T}d\tau\int_{-\infty}^{\infty}dx\,
\left(
-i\vec{\Pi}\dot{\vec{\phi}}+\CH(\vec{\phi},\vec{\phi}',\vec{\Pi})\right)
\label{2.2a}
\end{equation}
where $\CH$ is the Hamiltonian density (\ref{1.1a}) of the system.

If the kinetic term of a field theory Lagrangian is quadratic in the
time derivatives, e.\ g.\ for a field $w(\tau,x)$
\begin{equation}
\CL (w,w',\dot{w})=\frac12M(\dot{w})^2-{\cal U}(w,w')
\label{2.2b}
\end{equation}
(which means
$\vec{\pi}=M\dot{w}$), the momentum integration (which in this
case is a standard Gaussian integral) yields the well--known Feynman formula
\cite{Feynman}
\begin{equation}
{\cal Z} = \int {\cal D}\{w\}
e^{-\int_0^{\beta_T}d\tau\int_{-\infty}^{\infty}dx\,
\CL_ e(w,w',\dot{w})}
\label{2.2c}
\end{equation}
(${\cal D}\{\ldots\}$ denoting Lagrangian path measures which depend on
the mass $M$)
with the {\em Euclidean Lagrange density} 
\begin{equation}
\CL_ e(w,w',\dot{w})=\frac12M(\dot{w})^2+{\cal U}(w,w')
\label{2.2d}
\end{equation}
where dots now denote derivatives with respect to $\tau$.
Of course one can also compute the Euclidean Lagrangian of our Skyrme
model from its Minkowskian version (\ref{1}), 
\begin{equation}
\CL_ e(\vec{\phi},\vec{\phi}',\dot{\vec{\phi}}) =
\lambda_1\left\{\nu^2\left[(\dot{\vec{\phi}})^2(\vec{\phi}')^2 -
(\dot{\vec{\phi}}\cdot\vec{\phi}')^2 
\right]+\frac{1}{2}[(\dot{\vec{\phi}})^2 +
(\vec{\phi}')^2] + \mu^2 V(\phi^3)\right\}
\label{2.3}
\end{equation}
but we can not simply use it in formula (\ref{2.2a}) since
the $\dot{\vec{\phi}}$--dependence in (\ref{1}) 
is quadratic but involves a field--dependent mass factor, 
so that the momentum integration has 
to be analysed carefully as it will yield a measure factor \cite{Garrod}.

Here we perform a 
perturbative evaluation of ${\cal Z}$ around two extrema (there may 
certainly be more contributions): ${\cal Z}\approx {\cal Z}_0+{\cal Z}_1$.
${\cal Z}_0$ is evaluated around the vacuum $\vec{\varphi}_0=(0,0,1)$
and thus contains the usual perturbative vacuum fluctuations. We know that
this part has to be real.
${\cal Z}_1$ is evaluated around the sphaleron $\vec{\varphi}(x)$ which is
also an extremum of the Euclidean action $S_e[\vec{\phi}]$. We expect
this part to yield an imaginary contribution to the partition function since
we want to describe thermal transitions by the sphaleron. Assuming
$|{\cal Z}_1|\ll {\cal Z}_0$, we may write
\begin{equation}
\mbox{Im}(\beta_T F) = -\mbox{Im}(\ln {\cal Z}) \approx -\mbox{Im}\left[
\ln\left({\cal Z}_0\left(1+\frac{{\cal Z}_1}{{\cal Z}_0}\right)\right)\right]
\approx - \frac{1}{{\cal Z}_0}\mbox{Im}({\cal Z}_1).
\label{2.4}
\end{equation}

The evaluation around each of the two classical solutions
$\vec{\varphi}_i$, $i=0,1$ is done in a 
parametrization
similar to (\ref{1.4}), but now with spacetime dependent parameter fields
$\vec{f}(\tau,x)=\left(f^{(\parallel)}(\tau,x),f^{(\perp)}(\tau,x)\right)$:
\begin{equation}
\vec{\phi}(\tau,x) = \frac{1}{\sqrt{1+\left(f^{(\perp)}(\tau,x)\right)^2}}
\left(\begin{array}{c} 
\sin f^{(\parallel)}(\tau,x) \\ f^{(\perp)}(\tau,x) \\
\cos f^{(\parallel)}(\tau,x) \end{array} \right)
\label{2.4a}
\end{equation}
In terms of the
parameter fields $\vec{f}(\tau,x)$, the classical solutions 
$\vec{\varphi}_i(x)$
are given by $f^{(\parallel)}(\tau,x)=f_i(x)$, $f^{(\perp)}(\tau,x)=0$
with $f_0(x)=0$, $f_1(x)=f(x)=
4 \arctan\left(e^{\pm\frac{x-x_0}{\sqrt{\mu}}}\right)$. We add fluctuations
$\vec{v}(\tau,x)=\left(v^{(\parallel)}(\tau,x),v^{(\perp)}(\tau,x)\right)$ 
to both parameter fields to obtain  
the fluctuation ansatz
\begin{eqnarray}
& & \hspace*{-9mm}\vec{\phi}(\tau,x) \nonumber\\ 
& = & \frac{1}{\sqrt{1+\left(v^{(\perp)}(\tau,x)\right)^2}}
\left(\begin{array}{c} 
\sin\left(f_i(x)+v^{(\parallel)}(\tau,x)\right) \\ 
v^{(\perp)}(\tau,x) \\
\cos\left(f_i(x)+v^{(\parallel)}(\tau,x)\right)\end{array}\right) \label{2.5} \\
& \approx &
\left(\begin{array}{c} \sin f_i(x) + v^{(\parallel)}(\tau,x) \cos f_i(x) - 
\frac12 \left(
\left(v^{(\perp)}(\tau,x)\right)^2+\left(v^{(\parallel)}(\tau,x)\right)^2
\right) \sin f_i(x) \\
v^{(\perp)}(\tau,x) \\ \cos f_i(x) - v^{(\parallel)}(\tau,x) \sin f_i(x) - 
\frac12 \left(
\left(v^{(\perp)}(\tau,x)\right)^2+\left(v^{(\parallel)}(\tau,x)\right)^2
\right) \cos f_i(x) \end{array} \right) \nonumber \\
& & {}
+ \CO\left(\left(v^{(\parallel)},v^{(\perp)}\right)^3\right) \label{2.6} \\
& = & \vec{\varphi}_i(x) +
\left(\vec{\psi}\left(\vec{v})\right)\right)(\tau,x) + 
\CO\left(\left(v^{(\parallel)},v^{(\perp)}\right)^3\right)
\label{2.7}
\end{eqnarray}
(One should note 
that we are considering time dependent fluctuations!). We do not
attach indices to the fluctuations since 
it is clear that the fluctuations are different
for $\vec{\varphi}_0$ and $\vec{\varphi}_1$. 

Here,
\begin{eqnarray}
& & \hspace*{-16mm} 
\left(\vec{\psi}\left(\vec{v}\right)\right)(\tau,x) 
\nonumber \\
& = & \left(\begin{array}{c} v^{(\parallel)}(\tau,x) \cos f_i(x) - 
\frac12 \left(\left(v^{(\perp)}(\tau,x)\right)^2+
\left(v^{(\parallel)}(\tau,x)\right)^2\right) \sin f_i(x) \\
v^{(\perp)}(\tau,x) 
\\- v^{(\parallel)}(\tau,x) \sin f_i(x) - \frac12 \left(
\left(v^{(\perp)}(\tau,x)\right)^2+\left(v^{(\parallel)}(\tau,x)\right)^2\right)
\cos f_i(x) \end{array} \right) 
\label{2.8}
\end{eqnarray}
are the fluctuations around the solution ansatz 
$\vec{\varphi}_i(x)=(\sin f_i(x),0,\cos f_i(x))$, 
constructed such that they respect the sigma model constraint $\vec{\phi}^2=1$
(at least up two second order in 
$\vec{\psi}\left(\vec{v}\right)$). 
They are parametrized in terms of two fluctuation functions 
$\vec{v}(\tau,x)=\left(v^{(\perp)}(\tau,x),v^{(\parallel)}(\tau,x)\right)$ 
which represent
the two dimensions of a unit sphere.

We now insert (\ref{2.6}) into the Lagrangian 
(\ref{1}) and 
expand it to the mentioned second order in the
fluctuations which yields the linearized Lagrangian (around the classical 
solution $\vec{\varphi}_i$)
\begin{equation}
\tilde{\CL }_i(\vec{v},\vec{v}',\dot{\vec{v}}) = 
\lambda_1\left\{\frac12\left[
\left\{1+\nu^2
(f_i'(x))^2\right\}\left(\dot{v}^{(\perp)}\right)^2+
\left(\dot{v}^{(\parallel)}\right)^2\right] 
-{\cal U}_i\left(\vec{v},\vec{v}'\right)\right\}
\label{2.8a}
\end{equation}
where
\begin{eqnarray}
{\cal U}_i\left(\vec{v},\vec{v}'\right) & = & 
\frac12\left(\vec{v}'\right)^2
+ \left[\frac12(f'_i(x))^2+\mu^2V(\cos f_i(x))\right]\nonumber \\
& & 
+ {} \left[f_i'(x)\left(v^{(\parallel)}\right)'
-\mu^2V_{\phi^3}(\cos f_i(x))\sin f_i(x)\right]
\nonumber \\
& & - {} \frac12\left[(f_i'(x))^2+
\mu^2V_{\phi^3}(\cos f_i(x))\cos f_i(x)\right]\left(v^{(\perp)}\right)^2
\nonumber \\
& & - {} \frac{\mu^2}{2}\left[V_{\phi^3}(\cos f_i(x))\cos f_i(x)-
V_{\phi^3\phi^3}(\cos f_i(x))\sin^2 f_i(x)\right]
\left(v^{(\parallel)}\right)^2\nonumber \\
\label{2.8b}
\end{eqnarray}
where $V_{\phi^3}$ is the derivative of $V$ with respect to $\phi^3$.
Defining the linearized conjugate momenta
\begin{eqnarray}
\pi_{(\perp)} & := & \frac{\partial\tilde{\CL }}{\partial\dot{v}^{(\perp)}}=
\lambda_1k_i(x)\dot{v}^{(\perp)},\qquad k_i(x):=1+\nu^2
(f_i'(x))^2 \label{2.8c} \\
\pi_{(\parallel)} & := & 
\frac{\partial\tilde{\CL }}{\partial\dot{v}^{(\parallel)}}=
\lambda_1\dot{v}^{(\parallel)}
\label{2.8d}
\end{eqnarray}
it is now easy to write down the linearized Hamiltonian in terms of 
(parameter) fields and conjugate momenta 
$\vec{\pi}=\left(\pi_{(\parallel)},\pi_{(\perp)}\right)$, i.e.\
\begin{eqnarray}
\tilde\CH_i(\vec{v},\vec{v}',\vec{\pi}) & = & \vec{\pi}\cdot\dot{\vec{v}}-
\tilde\CL_ i(\vec{v},\vec{v}',\dot{\vec{v}})\nonumber \\
& = & \lambda_1\left\{\frac12\frac{\left(\pi_{(\perp)}\right)^2}{k_i(x)}
+\frac12\left(\pi_{(\parallel)}\right)^2+{\cal U}_i(u,v,u',v')\right\}
\label{2.8e}
\end{eqnarray}

To insert (\ref{2.8e}) into (\ref{2.2}), we also have to take care of 
the change in the integration measure. We are performing a substitution
$\vec{\phi}(\tau,x)\mapsto\vec{v}(\tau,x)$. Since we now use
parameter fields which respect the constraint $|\vec{\phi}|=1$, we get rid
of the delta distribution in the integral. From the general theory of 
distributions, it is well--known that for the parametrization 
$\vec{\phi}=\vec{\phi}(\vec{f})$,
\begin{eqnarray}
\prod_{i=1}^3[{\rm d}{\phi}^i(\tau,x)]
\delta(|\vec{\phi}|-1) & = & \left\|
\frac{\partial\vec{\phi}}{\partial f^{(\parallel)}}\wedge
\frac{\partial\vec{\phi}}{\partial f^{(\perp)}} \right\|
\left[{\rm d}f^{(\parallel)}(\tau,x)\right]
\left[{\rm d}f^{(\perp)}(\tau,x)\right] 
\nonumber \\
& = & \left(1+(f^{(\perp)}(\tau,x)^2)\right)^{-\frac32}
\left[{\rm d}f^{(\parallel)}(\tau,x)\right]
\left[{\rm d}f^{(\perp)}(\tau,x)\right]
\label{2.15a}
\end{eqnarray}
The change to the fluctuations as integration variables is simply a shift,
$f^{(\parallel)}(\tau,x)\mapsto v^{(\parallel)}(\tau,x)=
f^{(\parallel)}(\tau,x)-f_i(x)$, 
$f^{(\perp)}(\tau,x)\mapsto v^{(\perp)}(\tau,x)=f^{(\perp)}(\tau,x)$,
and we have
\begin{eqnarray}
& & \left(1+(f^{(\perp)}(\tau,x))^2\right)^{-\frac32}
\left[d f^{(\parallel)}(\tau,x)\right]
\left[d f^{(\perp)}(\tau,x)\right] \nonumber \\
& = & \left(1+\left(v^{(\perp)}(\tau,x)\right)^2\right)^{-\frac32}
\left[d v^{(\perp)}(\tau,x)\right]
\left[d v^{(\parallel)}(\tau,x)\right]\nonumber \\
& \approx & \left[d v^{(\perp)}(\tau,x)\right]
\left[d v^{(\parallel)}(\tau,x)\right],
\label{2.15b}
\end{eqnarray}
neglecting corrections of order 
$\CO\left(\left(v^{(\perp)}(\tau,x)\right)^2\right)$. We also neglect
corrections from the substitution 
$\vec{\Pi}(\tau,x)\mapsto\vec{\pi}(\tau,x)$ which 
is a more subtle point.

Therefore, we are now left with two contributions
to the partition function, given by the integrals ($i=0,1$)
\begin{eqnarray}
{\cal Z}_i & \approx & \int\left[d v^{(\perp)}(\tau,x)\right]
\left[d v^{(\parallel)}(\tau,x)\right]
\left[d \pi_{(\perp)}(\tau,x)\right]
\left[d \pi_{(\parallel)}(\tau,x)\right]
\nonumber \\
& & \qquad\quad \times 
e^{-\int_0^{\beta_T}d\tau\int_{-\infty}^{\infty}dx\,
\left(
-i\vec{\pi}\dot{\vec{v}}+\tilde\CH_i(\vec{v},\vec{v}',\vec{\pi})\right)}.
\label{2.15c}
\end{eqnarray}
The Hamiltonian in (\ref{2.15c}) is of standard type (\ref{2.2b}) with
an extra factor to the mass $M=\lambda_1$ (which reduces to $1$ for the
vacuum solution $\vec{\varphi}_0$ given by $f_0(x)=1$). Thus we may perform
the standard momentum integration (\ref{2.2c}) which yields
\begin{equation}
{\cal Z}_i\approx\int{\cal D}\left\{\sqrt{k_i(x)}v^{(\perp)}(\tau,x)\right\}
\int{\cal D}\left\{v^{(\parallel)}(\tau,x)\right\}
e^{-\tilde{S}_{e,i}[\vec{v}]}
\label{2.15d}
\end{equation}
where the Euclidean linearized action is given by (with partial integrations)
\begin{eqnarray}
\tilde{S}_{e,i}[\vec{v}] & = & \int_0^{\beta_T}d\tau\int_{-\infty}^{\infty}dx\,
\tilde\CL_{e,i}(\vec{v},\vec{v}',\dot{\vec{v}}) \nonumber \\
& = & \lambda_1\int_0^{\beta_T}d\tau\int_{-\infty}^{\infty}dx\,
\left(\frac12\left[
k_i(x)\left(\dot{v}^{(\perp)}\right)^2+
\left(\dot{v}^{(\parallel)}\right)^2\right] 
+{\cal U}_i(\vec{v},\vec{v}')\right)\nonumber \\
& = & S_e[\vec{\varphi}_i] \label{2.9} \\
& & {} - \lambda_1\int_0^{\beta_T}d\tau\int_{-\infty}^{\infty}dx\, 
v^{(\parallel)}(\tau,x)
\left[f''_i(x) + \mu^2 V_{\phi^3}(\cos f_i(x))\sin f_i(x)\right]
\nonumber \\
\label{2.10} \\ 
& & {} + \frac12 \lambda_1 \int_0^{\beta_T}d\tau\int_{-\infty}^{\infty}dx\,
 v^{(\parallel)}(\tau,x) \hat{{\cal G}}_{\parallel}^{(i)} 
v^{(\parallel)}(\tau,x) \nonumber \\
& & {} + \frac12 \lambda_1 \int_0^{\beta_T}d\tau\int_{-\infty}^{\infty}dx\,
v^{(\perp)}(\tau,x)  \hat{{\cal G}}_{\perp}^{(i)} v^{(\perp)}(\tau,x). 
\label{2.11}
\end{eqnarray}

For the vacuum we have $S_e[\vec{\varphi}_0] = 0$, and
\begin{equation}
S_e[\vec{\varphi}_1] = \beta_T\lambda_1
\int \left[\frac{1}{2} (f'(x))^2 + 
\mu^2V(\cos f(x)) \right] dx
\label{2.12}
\end{equation}
is the Euclidean action of the (static !) sphaleron 
which obviously equals its (static) energy (\ref{1.1}) times $\beta_T$,
$S_e[\vec{\varphi}_1]=S_{sph}(\beta_T)=\beta_T\CE[\vec{\varphi}_1]=
\beta_T\CE_{sph}$. 

The part linear in the  fluctuations, (\ref{2.10}), vanishes if
\begin{eqnarray}
& & f''_i(x) - \mu^2V_{\phi^3}(\cos f_i(x))\sin f_i(x)=0 \label{2.13allg} \\
& \Rightarrow & f''_i(x) - \mu^2\sin f_i(x) = 0, 
\qquad \mu^2 := \frac{\lambda_0}{\lambda_1}
\label{2.13}
\end{eqnarray}
after inserting the potential $V(\phi^3)=1-\phi^3$.
\footnote{
Eq.\ (\ref{2.13allg}) can also be solved for other potentials of the
class $V(\phi^3)=(1-\phi^3)^k$, $k\in I\!\!N$. For $k=2$, e.g., one
obtains $f(x) = 2\mbox{arccot}\left(\sqrt{2}\mu x\right)$.

Also, the fluctuation equations for these potentials can be obtained easily
from the general form of the equations given in the text. But already
for $k=2$, this yields eigenvalue equations to which no explicit solutions
are known.} This yields 
the vacuum ($i=0$) and the well--known classical static 
solution ($i=1$) (\ref{1.13}).

The quadratic
contributions (\ref{2.11}) are written in terms
of the (full, Euclidean) Gaussian fluctuation operators 
\begin{eqnarray}
\hat{{\cal G}}_{\parallel}^{(i)} 
& = & - \frac{\partial^2}{\partial \tau^2}
- \frac{\partial^2}{\partial x^2} -
\mu^2 \left\{V_{\phi^3}(\cos f_i(x))\cos f_i(x) -  
V_{\phi^3\phi^3}(\cos f_i(x))\sin^2 f_i(x) \right\} \nonumber \\ 
& = &  -  \frac{\partial^2}{\partial \tau^2}
- \frac{\partial^2}{\partial x^2} +
\mu^2\cos f_i(x)
\label{2.14} \\
\hat{{\cal G}}_{\perp}^{(i)} & = & - \{1 + \nu^2 (f'_i(x))^2\} 
\frac{\partial^2}{\partial \tau^2} - 
\frac{\partial^2}{\partial x^2} - 
\left\{\mu^2V_{\phi^3}(\cos f_i(x))\cos f_i(x) +
(f'_i(x))^2\right\}
\nonumber \\ 
& = &  - \{1 + \nu^2 (f'_i(x))^2\} 
\frac{\partial^2}{\partial \tau^2} - 
\frac{\partial^2}{\partial x^2} 
+\mu^2(3\cos f_i(x)-2)
\label{2.15} 
\end{eqnarray}
where we inserted $V(\phi^3)=1-\phi^3$.
These operators describe fluctuations parallel and perpendicular to
the solutions $\vec{\varphi}_i$.

Writing the definition of operator determinants \cite {determinants}
\begin{eqnarray}
\int{\cal D}\left\{v^{(\parallel)}(\tau,x)\right\}
e^{-\frac12\lambda_1\int_0^{\beta_T}d\tau\int_{-\infty}^{\infty}dx\ 
v^{(\parallel)}(\tau,x) 
\hat{{\cal G}}_{\parallel}^{(i)} v^{(\parallel)}(\tau,x)} 
& = & 
\frac{1}{\sqrt{\det\left(\hat{{\cal G}}_{\parallel}^{(i)}\right)}}, 
\label{2.19} \\
\int{\cal D}\left\{\sqrt{k_i(x)}v^{(\perp)}(\tau,x)\right\}
e^{-\frac12\lambda_1\int_0^{\beta_T}d\tau\int_{-\infty}^{\infty}dx\
v^{(\perp)}(\tau,x) \hat{{\cal G}}_{\perp}^{(i)} v^{(\perp)}(\tau,x)}
& = &
\frac{1}{\sqrt{\det\left(\hat{{\cal G}}_{\perp}^{(i)}\right)}}
\label{2.20}
\end{eqnarray}
(the ``mass'' $M=\lambda_1$ is absorbed in the path integration with
measure ${\cal D}\{\ldots\}$ depending on $\lambda_1$, and 
we will see why the measure factor $\sqrt{k_i(x)}$ in (\ref{2.20}) is
necessary for some similar reasons) yields in the one--loop approximation
\begin{equation}
{\cal Z}_i=\frac{e^{-S[\vec{\varphi}_i]}}
{\sqrt{\det\left(\hat{{\cal G}}_{\parallel}^{(i)}\right)
\det\left(\hat{{\cal G}}_{\perp}^{(i)}\right)}}.
\label{2.21}
\end{equation}
Inserting these results into (\ref{2.1}) finally leads to the following 
general formula for the transition rate:
\begin{equation}
\Gamma = -\frac{|\varrho|}{\pi}e^{-\beta_T\CE[\vec{\varphi}_1]}
\mbox{Im}\sqrt{\det\left(
\frac{(\hat{{\cal G}}_{\parallel}^{(0)}}{(\hat{{\cal G}}_{\parallel}^{(1)}}
\right)
\det\left(\frac{\hat{{\cal G}}_{\perp}^{(0)}}{\hat{{\cal G}}_{\perp}^{(1)}}
\right)}
\label{2.21b}
\end{equation}
We therefore have to evaluate the determinants of the Euclidean Gaussian 
fluctuation operators, using appropriate regularisation techniques.

\section{Fluctuations around the vacuum}

\label{sectionvacfluc}

We first consider the vacuum contribution ${\cal Z}_0$, i.\ e.\ $i=0$
in all
formulae of section \ref{section2}. This yields $S_e[\vec{\varphi}_0]=0$
and the measure factor $k_0(x)=1$, and the fluctuation operators are simply
\begin{equation}
\hat{{\cal G}}_{\perp}^{(0)}=\hat{{\cal G}}_{\parallel}^{(0)}=
\hat{{\cal G}}^{(0)}=- \frac{\partial^2}{\partial \tau^2}
- \frac{\partial^2}{\partial x^2} +
\mu^2=- \frac{\partial^2}{\partial \tau^2}
+\mu\left[\frac{\partial^2}{\partial y^2} + 1\right]
\label{vac1}
\end{equation}
with $y=\mu x$, $\mu>0$.
$\hat{{\cal G}}^{(0)}$ has only a continuous spectrum given by
\begin{equation}
\left(\frac{2\pi n}{\beta_T}\right)^2+\mu^2(1+k^2), \qquad k\in I\!\!R,
\label{vac2}
\end{equation}
the corresponding generalised eigenfunctions (which are not square integrable)
being $\Psi_{n,k}(\tau,x)=e^{\frac{2\pi in}{\beta_T}\tau}\psi_k(\mu x)$ with 
$\psi_k(y)=e^{iky}$.

Decomposing the fluctuations $v^{(\parallel)}$, 
$v^{(\perp)}$ in terms of the eigenfunctions 
$\Psi_{n,k}(\tau,x)$ (which is commonly called
``Fourier--Matsubara decomposition'' \cite{FouMats}) thus yields
\begin{equation}
\int{\cal D}\{v(\tau,x)\}
e^{-\frac12\lambda_1\int_0^{\beta_T}d\tau\int_{-\infty}^{\infty}dx\ 
v(\tau,x) \hat{{\cal G}}^{(0)} v(\tau,x)} 
= \prod_{k\in I\!\!R}\prod_{n=-\infty}^{\infty}
\left\{\left(\frac{2\pi n}{\beta_T}\right)^2+\mu^2(1+k^2)\right\}^{-\frac12}
\label{vac3}
\end{equation}
$v=v^{(\parallel)},v^{(\perp)}$.
This infinite product has to be regularised in a convenient way; usually
this is done by zeta function techniques \cite{determinants}. 
For fixed $k$, one can perform
the $n=-\infty\ldots\infty$--product which yields the partition function
of an harmonic oscillator with frequency $\omega(k):=\mu\sqrt{1+k^2}$,
\begin{eqnarray}
{\cal Z}(\omega(k)):=\frac{1}{2\sinh\frac{\omega(k)\beta_T}{2}} & = & 
\exp\left(-\beta_T\Phi(k^2;\beta_T,\mu)
\right),
\label{vac4}\\
\Phi(k^2;\beta_T,\mu) & := & \frac{\mu}{2}\sqrt{1+k^2}+
\frac{1}{\beta_T}\ln\left(1-e^{-\beta_T\mu\sqrt{1+k^2}}\right).
\label{vac5}
\end{eqnarray}
The notation $\Phi(k^2;\beta_T,\mu)=\Phi_1(k^2;\mu)+
\frac{1}{\beta_T}\Phi_2(k^2;\beta_T\mu)$ with
\begin{equation}
\Phi_1(k^2;\mu)=\frac{\mu}{2}\sqrt{1+k^2}, \qquad
\Phi_2(k^2;a)=\ln\left(1-e^{-a\sqrt{1+k^2}}\right)
\label{vac5a}
\end{equation}
will be useful in the following.

What remains to be done 
is the product over all $k\in I\!\!R$. Taking the logarithm of
(\ref{vac3}), we write it as an integral over $k$ weighted with the density
of the continuous vacuum fluctuation states $\rho_0$ 
(which will be discussed later). The final result
for the vacuum determinants is thus
\begin{equation}
\frac{1}{\sqrt{\det\left(\hat{{\cal G}}_{\parallel}^{(0)}\right)}}=
\frac{1}{\sqrt{\det\left(\hat{{\cal G}}_{\perp}^{(0)}\right)}}=
\exp\left(-\beta_T\int_{-\infty}^{\infty}dk\,
\rho_0(k)\Phi(k^2;\beta_T,\mu^2)\right)
\label{vac6}
\end{equation}

\section{Parallel fluctuation contributions to the transition rate}

\label{sectionparfluc}

The determinant of $\hat{{\cal G}}_{\parallel}^{(1)}$ can also be calculated
with a standard Fourier--Matsubara decomposition, i.e.\
\begin{eqnarray}
\hat{{\cal G}}_{\parallel}^{(1)} 
& = & -\frac{\partial^2}{\partial \tau^2}
-\frac{\partial^2}{\partial x^2} +
\mu^2\left\{1-2\mbox{sech}^2\left(\mu x\right)
\right\} \nonumber \\
& = & - \frac{\partial^2}{\partial \tau^2} + 
\hat{\CH}_{\parallel}
\label{2.16}
\end{eqnarray}
has eigenfunctions $\Psi_{nm}^{(\parallel)}(\tau,x)=
e^{\frac{2\pi in}{\beta_T}\tau}\psi_m^{(\parallel)}(x)$ with eigenvalues
\begin{equation}
\left(\frac{2\pi n}{\beta_T}\right)^2+\left(\omega_m^{(\parallel)}\right)^2
\label{2.16a}
\end{equation}
where $\left(\omega_m^{(\parallel)}\right)^2$ are the eigenvalues of
\begin{equation}
\hat{\CH}_{\parallel}\psi_m^{(\parallel)}=-
\frac{\partial^2\psi_m^{(\parallel)}}{\partial x^2} +
\mu^2\left\{1-2\mbox{sech}^2\left(\mu x\right)\right\}\psi_m^{(\parallel)}
=\left(\omega_m^{(\parallel)}\right)^2\psi_m^{(\parallel)}
\label{2.16b}
\end{equation}
which is just the static fluctuation equation (\ref{1.16}).
This is a special case of eq. (\ref{3.1}) discussed in Appendix \ref{app3}
with $U=s(s+1)\equiv 2$, $s=s_{\parallel}=1$
and $E=\frac{1}{\mu^2}\left(\omega_m^{(\parallel)}\right)^2-1$ which is known
as P\"oschl-Teller equation \cite{Schrodinger}. It has a normalised
zero mode 
$\psi_0^{(\parallel)}(x)=\sqrt{\frac{\mu}{2}}\mbox{sech}\left(\mu x\right)$, 
$\left(\omega_0^{(\parallel)}\right)^2=0$, 
and a continuum
$\left(\omega^{(\parallel)}\right)^2(k)=\mu^2(1+k^2)$. 
Decomposing the parallel fluctuations $v^{(\parallel)}$
into eigenstates of $\hat{{\cal G}}_{\parallel}^{(1)}$,
$v^{(\parallel)}(\tau,x)=\frac{1}{\sqrt{\beta_T}}
\sum_{n,m}c_{nm}^{(\parallel)}\Psi_{nm}^{(\parallel)}(\tau,x)$, yields
\begin{eqnarray}
& & \hspace*{-2cm} \int{\cal D}\left\{v^{(\parallel)}(\tau,x)\right\}
e^{-\frac{\lambda1}{2}\int_0^{\beta_T}d\tau\int_{-\infty}^{\infty}dx\ 
v^{(\parallel)}(\tau,x) 
\hat{{\cal G}}_{\parallel}^{(1)} v^{(\parallel)}(\tau,x)} \nonumber \\
& = & \prod_m\prod_{n=-\infty}^{\infty}
\left\{\left(\frac{2\pi n}{\beta_T}\right)^2+
\left(\omega_m^{(\parallel)}\right)^2\right\}^{-\frac12}
\label{2.16c}
\end{eqnarray}
where $m$ now runs over the values $m=0$ and $m=k$, $k\in I\!\!R$. Again
(assuming zeta function regularisation), the $n$--product yields one
harmonic oscillator for each fixed $m$. 

To illustrate this appearence of harmonic oscillators, one can also decompose
the parallel fluctuations in terms of the eigenfunctions of eq.\ (\ref{2.16b})
which can be normalized with respect to the scalar product
\begin{equation}
\langle\xi_1,\xi_2\rangle_{\parallel}
:=\int_{-\infty}^{\infty}\xi_1^*(x)\xi_2(x)dx
\label{2.16ca}
\end{equation}
on $L^2(I\!\!R)$: $\left\langle\psi_n^{(\parallel)},\psi_m^{(\parallel)}
\right\rangle_{\parallel}=\delta_{nm}$ (with the usual generalisation
for scattering states which we assume throughout in the following). This
decomposition 
\begin{equation}
v^{(\parallel)}(\tau,x)=\sum_m c_m^{(\parallel)}(\tau)\psi_m^{(\parallel)}(x)
\label{2.16caq}
\end{equation}
is related to the Fourier--Matsubara decomposition used above by
\begin{equation}
c_m^{(\parallel)}(\tau)=\frac{1}{\sqrt{\beta_T}}\sum_{n=-\infty}^{\infty}
c_{nm}^{(\parallel)}e^{\frac{2\pi i n}{\beta_T}\tau}.
\label{2.16caa}
\end{equation}
For fixed $\tau$, 
(\ref{2.16caq}) defines a mapping $L^2(I\!\!R)\rightarrow\ell$ which is an 
isometry with respect to the $L^2(I\!\!R)$ scalar product 
$\langle,\rangle_{\parallel}$. As this scalar product corresponds to the 
measure in the fluctuation path integral (\ref{2.16c}), we may use 
(\ref{2.16caq}) as substitution in the integral without unit Jacobian. This
yields
\begin{eqnarray}
& & \int{\cal D}\left\{v^{(\parallel)}(\tau,x)\right\}
e^{-\frac12\int_0^{\beta_T}d\tau\int_{-\infty}^{\infty}dx\ 
v^{(\parallel)}(\tau,x) 
\hat{{\cal G}}_{\parallel}^{(1)} v^{(\parallel)}(\tau,x)} 
\nonumber \\
& = & 
\prod_m\int {\cal D}\left\{c_m^{(\parallel)}(\tau)\right\}
e^{-\frac12 \int_0^{\beta_T}d\tau\left(\left|\dot{c}_m^{(\parallel)}(\tau)
\right|^2+\left(\omega_m^{(\parallel)}\right)^2\left|c_m^{(\parallel)}(\tau)
\right|^2\right)}
\nonumber \\
& = & \prod_m\frac{1}{2\sinh\frac{\omega_m^{(\parallel)}\beta_T}{2}}
\label{2.16cb}
\end{eqnarray}
Each $c_m^{(\parallel)}$--integral thus contributes one harmonic oscillator
with frequency $\omega_m^{(\parallel)}$. We will use this decomposition 
technique to evaluate the perpendicular fluctuation integral (\ref{2.20})
where the Fourier--Matsubara decomposition of the fluctuations is not so 
obvious. Details of the substitution used in (\ref{2.16cb}) can therefore be 
found in section \ref{section5}.

In (\ref{2.16c}), the only difference compared with the
vacuum calculation is the zero eigenvalue $m=0$ of $\CH_{\parallel}$. 
This yields a zero mode factor \cite{zero}
in the determinant of the parallel sphaleron fluctuations which is the
partition function of an harmonic oscillator with zero frequency, i.\ e.\ that 
of a free mode which has to be treated separately. This zero mode factor
originates
from the translation symmetry of the model which is broken by the classical
solution $\vec{\varphi}(x)$ which  is discussed in Appendix \ref{app2}.
The result of this (formal) evaluation of the $n$--product in (\ref{2.16c})
for $m=0$ is thus the replacement 
\begin{equation}
\prod_{n=-\infty}^{\infty}
\left\{\left(\frac{2\pi n}{\beta_T}\right)^2+
\left(\omega_0^{(\parallel)}\right)^2\right\}^{-\frac12}
=\frac{1}{2\sinh\frac{\omega_0{(\parallel)}\beta_T}{2}}\longrightarrow
L{\cal N}_{\parallel}:=2L\sqrt{\lambda_1}
\sqrt{\frac{\mu}{\pi\beta_T}}
\label{2.16d}
\end{equation}
where we introduced a finite space length $L$ to get rid of the volume 
divergence. (Later 
we will divide the transition rate by $L$ to obtain the transition rate per
unit volume which is independent of this procedure.)

The continuous product over $k\in I\!\!R$ is performed in exactly the same way
as in Section \ref{sectionvacfluc} and yields
\begin{equation}
\prod_{k\in I\!\!R}\frac{1}{2\sinh\frac{\omega^{(\parallel)}(k)\beta_T}{2}}
=\exp\left(-\beta_T\int_{-\infty}^{\infty}dk\,
\rho_{\parallel}(k)\Phi(k^2;\beta_T,\mu)\right)
\label{2.16e}
\end{equation}
where $\Phi(k^2;\beta_T\mu)$ 
is defined in (\ref{vac5}) and $\rho_{\parallel}(k)$
is the density of the continuum states of $\hat{\CH}_{\parallel}$.

Together with eq.\ (\ref{vac6}) we thus have
\begin{equation}
\sqrt{\det\left(
\frac{(\hat{{\cal G}}_{\parallel}^{(0)}}{(\hat{{\cal G}}_{\parallel}^{(1)}}
\right)} = L{\cal N}_{\parallel}
\exp\left(-\beta_T\int_{-\infty}^{\infty}dk\,
\left(\rho_{\parallel}(k)-\rho_0(k)\right)\Phi(k^2;\beta_T,\mu)\right).
\label{2.16f}
\end{equation}
This ``parallel fluctuation contribution'' to the transition
rate (\ref{2.21b}) is a real quantity; we thus expect the perpendicular
contribution to be imaginary.

The integral in (\ref{2.16f}), containing the difference of the densities of
continuum states of two Hamiltonians, can be evaluated using standard
techniques discussed in refs.\ 
\cite{tHooft,Bochka,MottWipf,Forgacs,Tsitsi,Carson}.
The main formula is
\begin{equation}
\int_{-\infty}^{\infty}dk\,
\left(\rho_{\parallel}(k)-\rho_0(k)\right)\Phi(k^2;\beta_T,\mu) = 
\frac{1}{2\pi}\int_{-\infty}^{\infty}dk\,\Phi(k^2;\beta_T,\mu)\frac{d}{dk}
\delta_{\parallel}(k)
\label{2.16g}
\end{equation}
where $\delta_{\parallel}(k)$ 
is the phase shift between incoming and outgoing asymptotic
plane waves in the scattering states of eq.\ (\ref{2.16b}) as discussed in 
Appendix \ref{app3}, eq.\ (\ref{3.27}): 
$\delta_{\parallel}(k)=\delta_1(k)=-2\arctan(k)$ since 
$s=s_{\parallel}=1$.

Inserting (\ref{2.16g}) into the integral of (\ref{2.16f})
yields for the continuum fluctuation contributions
\begin{equation}
\int_{-\infty}^{\infty}dk\,
\left(\rho(k)-\rho_0(k)\right)\Phi(k^2;\beta_T,\mu)
=-\frac{\mu}{2\pi}\int_{-\infty}^{\infty}dk\,\frac{dk}{\sqrt{1+k^2}}
+\frac{1}{\beta_T}h_{\parallel}(\beta_T\mu)
\label{2.16h}
\end{equation}
The first integral in (\ref{2.16h}) is logarithmically divergent; it is the sum 
of zero--point energy contributions of the oscillators in the Fourier--Matsubara
decomposition (\ref{2.16c}) and has to be renormalised. We ignore this here.

The function $h_{\parallel}$ is given by
\begin{eqnarray}
h_{\parallel}(a) & = & \frac{1}{2\pi} \int_{-\infty}^{\infty}dk\,\Phi_2(k;a)
\frac{d\delta_{\parallel}(k)}{dk} \nonumber \\
& = & -\frac{1}{\pi}\int_{-\infty}^{\infty}dk\,
\frac{\ln\left(1-e^{-a\sqrt{1+k^2}}\right)}{1+k^2} >0.
\label{2.16i}
\end{eqnarray}
The integral is convergent for all $a>0$, and
\begin{equation}
h_{\parallel}(\beta_T\mu)\sim -\ln(\beta_T\mu)>0
\label{2.16j}
\end{equation}
in the high temperature limit 
$\beta_T\rightarrow 0$ ($a=\beta_T\mu\rightarrow 0$) which is shown in 
Appendix \ref{app4}, eq.\ (\ref{h6}).

Finally, collecting all results, the parallel fluctuation contribution
to the transition rate (\ref{2.16f}) reads
\begin{equation}
\sqrt{\det\left(
\frac{(\hat{{\cal G}}_{\parallel}^{(0)}}{(\hat{{\cal G}}_{\parallel}^{(1)}}
\right)} = L{\cal N}_{\parallel}e^{-h_{\parallel}(\beta_T\mu)}.
\label{2.16k}
\end{equation}

\section{Perpendicular fluctuation contributions to the transition rate}

\label{section5}

In order to calculate the remaining perpendicular fluctuation contribution to
the transition rate (\ref{2.21b}), we have to evaluate the perpendicular 
fluctuation integral (\ref{2.20})
($i=1$) around the sphaleron. Here the measure factor (\ref{2.8c}),
\begin{equation}
k_1(x)\equiv k(x):=1+\nu^2(f'(x))^2=1+4\kappa^2\mbox{sech}^2\left(\mu x\right)
\label{5.0}
\end{equation}
with $\kappa^2=\mu^2\nu^2$ 
is the important new ingredient which contains
the influence of the Skyrme term on the model. The fluctuation integral reads
\begin{equation}
\int{\cal D}\left\{\sqrt{k(x)}v^{(\perp)}(\tau,x)\right\}
e^{-\frac12\lambda_1\int_0^{\beta_T}d\tau\int_{-\infty}^{\infty}dx\
v^{(\perp)}(\tau,x) \hat{{\cal G}}_{\perp}^{(1)} v^{(\perp)}(\tau,x)}
\label{5.1}
\end{equation}
with
\begin{equation}
\hat{{\cal G}}_{\perp} = - k(x)
\frac{\partial^2}{\partial \tau^2}
+ \hat{\CH}_{\perp}, \qquad
\hat{\CH}_{\perp}
= - \frac{\partial^2}{\partial x^2} +
\mu^2 \left\{1-6\mbox{sech}^2\left(\mu x\right)\right\}
\label{2.17}
\end{equation}
where $\hat{\CH}_{\perp}$ is 
the static Gaussian fluctuation operator (\ref{1.17})
of the static energy functional.

\subsection{The measure--modified harmonic oscillator decomposition}

To obtain the desired 
Fourier--Matsubara decomposition in (\ref{5.1}), we solve the
following ``measure--modified static fluctuation equation'' \cite{measure}
\begin{equation}
\hat{\CH}_{\perp}\psi^{(\perp)}(x)=\left(\omega^{(\perp)}\right)^2
k(x)\psi^{(\perp)}(x)
\label{5.3}
\end{equation}
which is a general Sturm--Liouville problem on $L^2(I\!\!R)$ 
with the metric factor $k(x)$ on the right hand side
multiplying the eigenvalue \cite{SturmLiou}. 
(This equation is different from (\ref{1.17}).
We are no longer considering the sphaleron in the space of static field
configurations where it was first constructed, cf.\ Section \ref{section1}, 
but embedded it into 
the full euclidean field theory!)
From Sturm--Liouville theory, we know that the eigenfunctions of (\ref{5.3})
are orthogonal with respect to the scalar product
\begin{equation}
\langle \xi_1,\xi_2 \rangle_{\perp} := \int_{-\infty}^{\infty}
\xi_1^*(x)\xi_2(x) k(x) dx.
\label{5.5}
\end{equation}
We may thus orthonormalize the eigenfunctions 
$\hat{\CH}_{\perp}\psi_m^{(\perp)}=\left(\omega_m^{(\perp)}\right)^2
k(x)\psi_m^{(\perp)}$ of (\ref{5.3}),
$\langle \psi_m^{(\perp)},\psi_l^{(\perp)}\rangle_{\perp} = \delta_{lm}$
and expand the fluctuations $v^{(\perp)}$ in a Fourier--series with 
$\tau$--dependent periodic coefficients:
\begin{equation}
v^{(\perp)}(\tau,x) = \sum_m c_m^{(\perp)}(\tau)\psi_m^{(\perp)}(x)
\label{5.6}
\end{equation}
(This expansion again contains generalised scattering eigenfunctions in the 
usual sense).

The fluctuations are real, so that 
$\sum_m c_m^{(\perp)}(\tau)\psi_m^{(\perp)}(x) = 
\sum_m \left(c_m^{(\perp)}\right)^*(\tau)\left(\psi_m^{(\perp)}\right)^*(x)$. 
Inserting
(\ref{5.6}) into the integrand of (\ref{5.1}) yields
\begin{eqnarray}
& & \hspace*{-10mm} v^{(\perp)}(\tau,x)
\hat{{\cal G}}_{\perp}^{(1)}v^{(\perp)}(\tau,x) \nonumber \\
& = &  
\left(\sum_m \left(c_m^{(\perp)}\right)^*(\tau)
\left(\psi_m^{(\perp)}\right)^*(x) \right)
\left[ -k(x)\frac{\partial^2}{\partial\tau^2}+
\hat{\CH}_{\perp}\right]
\left( \sum_l c_l^{(\perp)}(\tau)\psi_l^{(\perp)}(x)\right) \nonumber \\
& = & 
\left(\sum_m  \left(c_m^{(\perp)}\right)^*(\tau)
\left(\psi_m^{(\perp)}\right)^*(x) \right)\nonumber \\
& & \qquad\qquad \times
\left( \sum_l \left[ -k(x)\ddot{c}_l^{(\perp)}(\tau)\psi_l^{(\perp)}(x)+
\left(\omega_l^{(\perp)}\right)^2
k(x)c_l^{(\perp)}(\tau)\psi_l^{(\perp)}(x)\right]\right) \nonumber \\
& = & \sum_m \sum_l 
\left[-\left(c_m^{(\perp)}\right)^*(\tau)
\ddot{c}_l^{(\perp)}(\tau)+\left(\omega_l^{(\perp)}\right)^2
\left(c_m^{(\perp)}\right)^*(\tau)c_l^{(\perp)}(\tau)\right]\nonumber \\
& & \qquad\qquad\times\left[
\left(\psi_m^{(\perp)}\right)^*(x)
\psi_l^{(\perp)}(x)k(x)dx\right],
\label{5.6a}
\end{eqnarray}
thus
\begin{eqnarray}
& &  \hspace*{-8mm}
\int_0^{\beta_T}d\tau\int_{-\infty}^{\infty}dx\, v^{(\perp)}(\tau,x)
\hat{{\cal G}}_{\perp}^{(1)}v^{(\perp)}(\tau,x) \nonumber \\
& = & \sum_m \sum_l 
\left[\int_0^{\beta_T}d\tau\left(-\left(c_m^{(\perp)}\right)^*(\tau)
\ddot{c}_l^{(\perp)}(\tau)+\left(\omega_l^{(\perp)}\right)^2
\left(c_m^{(\perp)}\right)^*(\tau)c_l^{(\perp)}(\tau)\right)\right]\nonumber \\
& & \qquad\qquad\times
\underbrace{\langle \psi_m^{(\perp)},\psi_l^{(\perp)}\rangle_{\perp}
}_{=\delta_{lm}} 
\nonumber \\
& = & \sum_m 
\int_0^{\beta_T}d\tau\left(-\left(c_m^{(\perp)}\right)^*(\tau)
\ddot{c}_m^{(\perp)}(\tau)+\left(\omega_m^{(\perp)}\right)^2
\left(c_m^{(\perp)}\right)^*(\tau)c_m^{(\perp)}(\tau)\right)
\nonumber \\
& = & \sum_m 
\int_0^{\beta_T}d\tau\left(\left|\dot{c}_m^{(\perp)}(\tau)\right|^2+
\left(\omega_m^{(\perp)}\right)^2\left|c_m^{(\perp)}(\tau)\right|^2\right).
\label{5.7}
\end{eqnarray}
This is a sum over harmonic oscillators. 
In particular, there is no contribution from the measure factor $k$ in 
this sum (it disappears with the help of Sturm--Liouville theory).

The mapping $v^{(\perp)}\mapsto c_m^{(\perp)}$ for fixed $\tau$ is an
isometry $L^2(I\!\!R)\rightarrow\ell^2$ with respect to the scalar product
$\langle,\rangle_{\perp}$ containing the measure factor $k(x)$ which also 
appears in the path integral measure, so we may substitute (\ref{5.6}) into
(\ref{5.1}) with unit Jacobian:
\begin{eqnarray}
& & \hspace*{-6mm} \int{\cal D}\left\{\sqrt{k(x)}v^{(\perp)}(\tau,x)\right\}
e^{-\frac12\lambda_1\int_0^{\beta_T}d\tau\int_{-\infty}^{\infty}dx\
v^{(\perp)}(\tau,x) \hat{{\cal G}}_{\perp}^{(1)} v^{(\perp)}(\tau,x)}
\nonumber \\
& = & \prod_m\int{\cal D}\left\{c_m^{(\perp)}(\tau)\right\}
e^{-\frac12\int_0^{\beta_T}d\tau\left(\left|\dot{c}_m^{(\perp)}(\tau)\right|^2+
\left(\omega_m^{(\perp)}\right)^2\left|c_m^{(\perp)}(\tau)\right|^2\right)}
\label{5.7aa}
\end{eqnarray}

Now it is easy to obtain the Fourier--Matsubara decomposition
of the perpendicular fluctuations. With a Fourier expansion of the coefficients
$c_m^{(\perp)}(\tau)$,
\begin{equation}
c_m^{(\perp)}(\tau)=\frac{1}{\sqrt{\beta_T}}
\sum_{n=-\infty}^{\infty}c_{nm}^{(\perp)}e^{\frac{2\pi in}{\beta_T}\tau},
\label{5.7a}
\end{equation}
we have 
\begin{equation}
v^{(\perp)}(\tau,x) = \frac{1}{\sqrt{\beta_T}}
\sum_{n,m} c_{nm}^{(\perp)}e^{\frac{2\pi in}{\beta_T}\tau}\psi_m^{(\perp)}(x)
\label{5.7b}
\end{equation}
similar to that of the parallel fluctuations. In terms of the 
$c_{nm}^{(\perp)}$,
the operator $\hat{{\cal G}}_{\perp}^{(1)}$ is completely diagonalised,
\begin{equation}
\int_0^{\beta_T}d\tau\int_{-\infty}^{\infty}dx\, v^{(\perp)}(\tau,x)
\hat{{\cal G}}_{\perp}^{(1)}v^{(\perp)}(\tau,x)=\sum_{n,m}c_{nm}^{(\perp)}
\left[\left(\frac{2\pi n}{\beta_T}\right)^2+\left(\omega_m^{(\perp)}\right)^2
\right]\left(c_{nm}^{(\perp)}\right)^*,
\label{5.7c}
\end{equation}
and the substitution $v^{(\perp)}(\tau,x)\mapsto c_{nm}^{(\perp)}$
is again an isometry with respect to the path integral measure, thus
\begin{eqnarray}
& & \hspace*{-10mm} \int{\cal D}\left\{\sqrt{k(x)}v^{(\perp)}(\tau,x)\right\}
e^{-\frac12\lambda_1\int_0^{\beta_T}d\tau\int_{-\infty}^{\infty}dx\
v^{(\perp)}(\tau,x) \hat{{\cal G}}_{\perp}^{(1)} v^{(\perp)}(\tau,x)}
\nonumber \\
& = & \int \prod_m\prod_{n=-\infty}^{\infty} 
{\cal D}\left\{c_{nm}^{(\perp)}\right\} 
e^{-\frac12
\sum_{n,m}c_{nm}^{(\perp)}
\left[\left(\frac{2\pi n}{\beta_T}\right)^2+
\left(\omega_m^{(\perp)}\right)^2\right]
\left(c_{nm}^{(\perp)}\right)^*}
\nonumber \\
& = & \prod_m\prod_{n=-\infty}^{\infty}
\left\{\left(\frac{2\pi n}{\beta_T}\right)^2+
\left(\omega_m^{(\perp)}\right)^2\right\}^{-\frac12}.
\label{5.7d}
\end{eqnarray}
with unit Jacobian.

Assuming zeta function regularisation \cite{determinants}, 
we first perform the $n$--product
to obtain
\begin{equation}
\int{\cal D}\left\{\sqrt{k(x)}v^{(\perp)}(\tau,x)\right\}
e^{-\frac12\lambda_1\int_0^{\beta_T}d\tau\int_{-\infty}^{\infty}dx\
v^{(\perp)}(\tau,x) \hat{{\cal G}}_{\perp}^{(1)} v^{(\perp)}(\tau,x)} = 
\prod_m \frac{1}{2\sinh\frac{\omega_m^{(\perp)}\beta_T}{2}}.
\label{5.7e}
\end{equation}
This result is also clear from (\ref{5.7aa}), but to controll the necessary
renormalisation steps (which is not our aim here), one has to return to the
Fourier--Matsubara decomposition.

\subsection{Solution and discussion of the measure--modified
static fluctuation equation}

In order to use the harmonic oscillator decompostion (\ref{5.7e}),
we have to solve (\ref{5.3}).
This Sturm--Liouville equation may be rewritten as (with 
$y=\mu x$)
\begin{equation}
\left\{\frac{d^2}{dy^2}+\left[\frac{\left(\omega^{(\perp)}\right)^2}{\mu^2}
-1\right]+\left[6+4\nu^2\left(\omega^{(\perp)}\right)^2\right]
\sech^2y\right\}\psi^{(\perp)}(y)=0.
\label{5.9}
\end{equation}
which is eq.\ (\ref{3.1}) with
\begin{eqnarray}
E & = & E\left(\left(\omega^{(\perp)}\right)^2\right)
=\frac{\left(\omega^{(\perp)}\right)^2}{\mu^2}-1 \quad \Longrightarrow \quad
\epsilon = \sqrt{1-
\frac{\left(\omega^{(\perp)}\right)^2}{\mu^2}} \label{5.10a} \\
U & = & U\left(\left(\omega^{(\perp)}\right)^2\right)
=6+4\nu^2\left(\omega^{(\perp)}\right)^2 \quad \Longrightarrow \quad
s=\sqrt{\frac{25}{4}+4\nu^2\left(\omega^{(\perp)}\right)^2}-\frac12.
\label{5.10}
\end{eqnarray}
It is convenient to discuss the results in terms of 
$\tilde{\omega}^2:=\frac{1}{\mu^2}\left(\omega^{(\perp)}\right)^2$ 
where the constant 
$\mu^2=\frac{\lambda_0}{\lambda_1}$ does not depend on the Skyrme coupling
constant $\lambda_2$, so that it is possible to discuss the Skyrme--less limit
$\lambda_2\rightarrow 0 $ in terms of $\tilde{\omega}^2$ which will depend
only on the one parameter 
$\kappa^2=\mu^2\nu^2=\frac{\lambda_0\lambda_2}{\lambda_1^2}$.

There are discrete eigenvalues
$\tilde{\omega}^2_m$ for all $m\in I\!\!N_0$ with
\begin{equation}
T_{U\left(\left(\omega_m^{(\perp)}\right)^2\right)}(m)=
\sqrt{\frac{25}{4}+4\kappa^2\tilde{\omega}_m^2}-\left(m+\frac12\right)>0
\label{5.11}
\end{equation}
and 
$U\left(\left(\omega^{(\perp)}\right)^2\right)=6+4\kappa^2\tilde{\omega}^2>0$ 
which are
conditions for the existence of an eigenvalue which depend on the eigenvalue 
itself. If a discrete eigenvalue exists, it is given by (\ref{3.13})
which means it is a solution of
\begin{equation}
-E\equiv 1-\tilde{\omega}_m^2
=\left[\sqrt{\frac{25}{4}+4\kappa^2\tilde{\omega}_m^2}-\left(m+\frac12\right)
\right]^2.
\label{5.12}
\end{equation}
We first prove that there are no solutions of (\ref{5.12}) with $m>2$ which
satisfy (\ref{5.11}). Assume that such a solution $\tilde{\omega}_m^2$ ($m>2$)
exists. If it is positive, $\tilde{\omega}_m^2>0$, then 
$1-\tilde{\omega}_m^2<1$, but the right hand side of eq.\ (\ref{5.12}) may be 
estimated from below:
\begin{equation}
\left[\sqrt{\frac{25}{4}+4\kappa^2\tilde{\omega}_m^2}-\left(m+\frac12\right)
\right]^2\ge\left[\sqrt{\frac{25}{4}}-\left(m+\frac12\right)\right]^2
=[2-m]^2\ge1 
\label{5.13}
\end{equation}
for $m>2$. So $\tilde{\omega}_m^2$ has to be negative or zero,
$\tilde{\omega}_m^2<0$, contrary to the assumption. 
If $\tilde{\omega}_m^2<0$ is so negative that
$6+4\kappa^2\tilde{\omega}^2<0$, it cannot be a discrete eigenvalue because
then $U(\omega_m^2)<0$. Otherwise, we may estimate (\ref{5.11}) from above:
\begin{equation}
\sqrt{\frac{25}{4}+4\kappa^2\tilde{\omega}_m^2}-\left(m+\frac12\right)
\le\sqrt{\frac{25}{4}}-\left(m+\frac12\right)=2-m<0
\label{5.14}
\end{equation}
for $m>2$. Therefore, a solution of (\ref{5.11}) and (\ref{5.12})
cannot exist for $m>2$. 

Next, we solve (\ref{5.12}) naively by squaring the square root 
(we have to check whether the solutions we obtain really solve 
(\ref{5.12})) to get a quadratic equation
\begin{equation}
A(\kappa^2)(\tilde{\omega}^2)^2+B(m,\kappa^2)\tilde{\omega}^2+C(m)=0
\label{5.15}
\end{equation}
with
\begin{eqnarray}
A(\kappa^2) & = & (1+4\kappa^2)^2 \nonumber \\
B(m,\kappa^2) & = & \frac{21}{4}(1+4\kappa^2)+2\left(m+\frac12\right)^2
(1-4\kappa^2)  \nonumber \\
C(m) & = & \left(\frac{25}{4}-\left(m+\frac12\right)^2\right)^2
-2\left(\frac{25}{4}+\left(m+\frac12\right)^2\right)+1
\label{5.16}
\end{eqnarray}
The general solution of (\ref{5.15}) is
\begin{equation}
\tilde{\omega}^2 = \frac{-B(m,\kappa^2)\pm\sqrt{\Delta(m,\kappa^2)}}{2A(\kappa^2)}
\label{5.17}
\end{equation}
where $\Delta(m,\kappa^2)$ is the discriminant
\begin{equation}
\Delta(m,\kappa^2) = 4\left(m+\frac12\right)^2
\left[25+4\left(29-4\left(m+\frac12\right)^2\right)\kappa^2+64
(\kappa^2)^2\right]
\label{5.18}
\end{equation}
It is easy to check that $\Delta(m,\kappa^2)$ is nonnegative if
$\kappa^2\ge 0$ for $m=0,1,2$. 

Next, we insert the solutions (\ref{5.17}),
i.\ e.\ 
\begin{eqnarray}
\tilde{\omega}^2_{m,\pm} & = & 
-\left[\frac{21}{4(1+4\kappa^2)}+\left(m+\frac12\right)^2
\frac{(1-4\kappa^2)}{(1+4\kappa^2)^2}\right] \nonumber \\
& & \qquad\quad {} \pm
\frac{\left(m+\frac12\right)}{(1+4\kappa^2)^2}
\sqrt{25+4\left(29-4\left(m+\frac12\right)^2
\right)\kappa^2+64(\kappa^2)^2}
\label{5.19}
\end{eqnarray}
into (\ref{5.12}) and (\ref{5.11}) to determine the allowed signs 
(we already remarked that we have to check the solutions). We have
\begin{equation}
\sqrt{\frac{25}{4}+4\kappa^2\tilde{\omega}_{m,\pm}^2} = 
\frac{\left|8\left(m+\frac12\right)\pm\sqrt{25+
4\left(29-4\left(m+\frac12\right)^2
\right)\kappa^2+64(\kappa^2)^2}\right|}
{2(1+4\kappa^2)}.
\label{5.20}
\end{equation}
The modulus is always positive for the upper (plus) sign, and one can easily
check that this is a solution of eq.\ (\ref{5.12}) which satisfies
(\ref{5.11}) for all $\kappa^2\ge 0$. But for the lower (minus) sign, we 
have to distinguish two cases: For 
$0\le\kappa^2 <\frac{25}{16}\frac{1}{\left(m+\frac12\right)^2-1}$, 
$\tilde{\omega}^2_{m,-}$ does not solve (\ref{5.12}). Otherwise, 
for $\kappa^2\ge \frac{25}{16}\frac{1}{\left(m+\frac12\right)^2-1}$, it solves
this equation, but the condition (\ref{5.11}) is not fulfilled. We therefore
have to reject $\tilde{\omega}^2_{m,-}$ as solution to (\ref{5.12}) under
the condition (\ref{5.11}). So finally, there are three discrete eigenvalues
of (\ref{5.9}) given by
\begin{eqnarray}
\omega_-^2(\kappa^2)=\tilde{\omega}^2_0(\kappa^2) = 
\frac{\left(\omega_0^{(\perp)}\right)^2}{\mu^2} 
& = & \frac{-\left(20\kappa^2+\frac{11}{2}\right)+
\frac12\sqrt{25+112\kappa^2+64(\kappa^2)^2}}{(1+4\kappa^2)^2} <0 
\nonumber \\
\label{5.21} \\
\tilde{\omega}^2_1(\kappa^2) = 
\frac{\left(\omega_1^{(\perp)}\right)^2}{\mu^2} & = & 0 \label{5.22} \\
\omega_+^2(\kappa^2)=\tilde{\omega}^2_2(\kappa^2) = 
\frac{\left(\omega_2^{(\perp)}\right)^2}{\mu^2} 
& = & \frac{4\kappa^2-\frac{23}{2}+
\frac52\sqrt{25+16\kappa^2+64(\kappa^2)^2}}{(1+4\kappa^2)^2} >0. \label{5.23}
\end{eqnarray}
There is a zero eigenvalue $\left(\omega_1^{(\perp)}\right)^2=0$.
In Appendix \ref{app2}, we show that this zero mode is due to the
$U(1)$ rotation symmetry of the Euclidean action (\ref{2.12}) which is broken 
by the classical sphaleron solution $\vec{\varphi}_1(x)$ (analogous to the
translation zero mode which can be found in the spectrum of the
parallel fluctuations). This effect is independent of $\kappa^2$, so the
zero mode exists for all $\kappa^2$.

\begin{figure}
\begin{center}
{\includegraphics[scale=0.9]{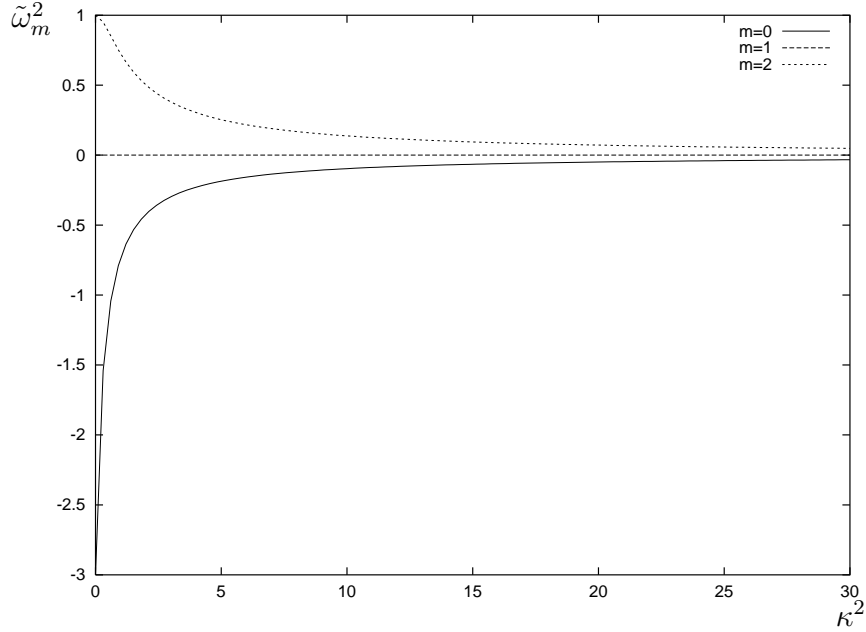}}
\put(-320,220){\makebox(0,0){$\tilde{\omega}_m^2$}}
\put(-10,-5){\makebox(0,0){$\kappa^2$}}
\end{center}
\caption{Discrete eigenvalues $\tilde{\omega}_m^2(\kappa^2)$ of
$\hat{{\cal G}}_{\perp}$ } \vspace{6mm}
\label{fig1}
\end{figure}

It is interesting to analyse the $\kappa^2$--dependence of the other two 
discrete eigenvalues which is shown in Fig.\ \ref{fig1}. For 
$\kappa^2\rightarrow 0$, we have $\tilde{\omega}^2_0(\kappa^2=0)=-3$ which
is the Skyrme--less result ($\lambda_2\rightarrow 0$). In the same
limit, we have $\tilde{\omega}^2_2(\kappa^2=0)=1$ which explains why
there are only two discrete eigenvalues in the original model \cite{MottWipf}:
Without a Skyrme term, the third eigenvalue is the lower bound of the
continuum. In the limit $\kappa^2\rightarrow \infty$, all three eigenvalues
tend to zero. This can be interpreted as the $\lambda_1\rightarrow 0$ limit
of the model in which there is no sphaleron solution (and therefore no 
stability spectrum).

The corresponding eigenfunctions are given by (\ref{3.14a}) were $s$ now 
depends on $\left(\omega_m^{(\perp)}\right)^2$ 
(cf.\ \ref{5.10}). Using (\ref{5.20}), we have
\begin{equation}
s=s_m(\kappa^2):=\frac{8m\kappa^2-1+\sqrt{25+4\left(29-4\left(m+\frac12\right)^2
\right)\kappa^2+64(\kappa^2)^2}}{2(1+4\kappa^2)}.
\label{5.24}
\end{equation}
For $m=1$, this yields $s_1(\kappa^2)=2$ which does not depend on $\kappa^2$.
The explicit form of the eigenfunctions is thus 
\begin{eqnarray}
\psi_0^{(\perp)}(y) & = & \psi_0^{(\perp)}(y;\kappa^2) 
= N_0^{(\perp)}\mbox{sech}^{s_0(\kappa^2)}y
\label{5.25} \\
\psi_1^{(\perp)}(y) & = & N_1^{(\perp)}\mbox{sech} y \tanh y \label{5.26} \\
\psi_2^{(\perp)}(y) & = & \psi_2^{(\perp)}(y;\kappa^2) =
N_2^{(\perp)}\mbox{sech}^{(s_2(\kappa^2)-2)}(y)
\left(1-\frac{2s_2(\kappa^2)-1}{2s_2(\kappa^2)-2}
\mbox{sech}^2 y\right) \label{5.27}
\end{eqnarray}
where $N_i^{(\perp)}$ are appropriate normalisation constants.
Like the zero eigenvalue, the zero mode (\ref{5.26}) is 
independent of $\kappa^2$
and agrees with that given in the literature \cite{MottWipf}. In the limit 
$\kappa^2\rightarrow 0$, we
have $s_0(0)=2$, so $\psi_0^{(\perp)}(y;\kappa^2=0)=\mbox{sech}^2y$ 
which again is
Mottola--Wipf's result. The $\kappa^2\rightarrow 0$ limit of $\psi_2^{(\perp)}$
is even
more interesting, because there is no such third discrete eigenvalue in the
Skyrme--less theory. In fact, we have $s_2(0)=2$ and therefore
\begin{equation}
\lim_{|y|\rightarrow \infty} \lim_{\kappa^2\rightarrow 0}
\psi_2^{(\perp)}(y;\kappa^2) = 
1 \quad \Longrightarrow \quad \psi_2^{(\perp)}(\cdot;\kappa^2=0)
\not\in L^2(I\!\!R).
\label{5.28}
\end{equation}
But since $s_2(\kappa^2)\ge 2$ for $\kappa^2\ge 0$ and 
$\psi_2^{(\perp)}(y;\kappa^2)
\stackrel{|y|\rightarrow
\infty, \kappa^2>0}{\longrightarrow} 0 $
exponentially, $\psi_2^{(\perp)}$ 
is a square integrable eigenfunction (and $\omega_2^2$
really a discrete eigenvalue) for $\kappa^2>0$. 

The continuous spectrum starts from $E>0$, i.\ e.\ 
$\omega^2(k)>\mu^2$. 
We denote the continuous eigenvalues
by $\omega^2(k)=\mu^2(1+k^2)$, $k\in I\!\!R$. 
The asymptotic behaviour of the scattering states belonging to $\omega^2(k)$
is discussed in Appendix \ref{app3}, (\ref{3.23}--\ref{3.25}), 
Now the parameter $s$ is not
fixed but depends on $k$ (and $\kappa^2$):
\begin{equation}
s=s_{\perp}(k;\kappa^2)=\sqrt{\frac{25}{4}+4\kappa^2(1+k^2)}-\frac12\in I\!\!R,
\label{5.29}
\end{equation}
and the coefficient of the incoming wave (\ref{3.25}) is
\begin{equation}
B(k;\kappa^2)=
\frac{\Gamma(-ik)\Gamma(1-ik)}{\Gamma(-ik-s(k;\kappa^2))
\Gamma(-ik+s(k;\kappa^2)+1)}.
\label{5.30}
\end{equation}

\subsection{Evaluation of the perpendicular fluctuation determinant}

With the above results we perform the discrete part of the product 
(\ref{5.7e}). From $m=2$, we simply get a harmonic oscillator factor
\begin{equation}
\frac{1}{2\sinh\left(\frac{\omega_+(\kappa^2)\beta_T\mu}{2}\right)}.
\label{5.28a}
\end{equation}
Next, we consider $m=0$. For sufficiently large $T$, there is only one negative
factor in the product (\ref{5.7d}) which means that ${\cal G}_{\perp}^{(1)}$
has exactly one negative eigenvalue. Then we may interpret the negative mode
as giving rise to an imaginary factor in ${\cal Z}_1$,
according to the 
prescription
\begin{equation}
\frac{1}{2\sinh\left(\frac{\omega_-(\kappa^2)\beta_T\mu}{2}\right)} 
\longrightarrow
\frac12\frac{1}{2i\sin\left(\frac{|\omega_-(\kappa^2)|\beta_T\mu}{2}\right)}.
\label{5.28b}
\end{equation}
where the additional factor $\frac12$ arises from the distortion of the
non--Gaussian contour over half of its range \cite{Coleman}.

Finally, $m=0$ in (\ref{5.7d}) is the zero mode contribution 
of $\tilde{\omega}_1^2=0$ which is discussed in Appendix
\ref{app2}, yielding (\ref{zero10})
\begin{equation}
\frac{1}{2\sinh\frac{\tilde{\omega}_1\beta_T}{2}}\longrightarrow
{\cal N}_{\perp}:=4\sqrt{\lambda_1}
\sqrt{\frac{\pi}{3\beta_T\mu}}\sqrt{1+\frac85\kappa^2}
\label{5.28c}
\end{equation}
(since this zero mode results from the $U(2)$--rotation symmetry of the 
Lagrangian broken by the sphaleron, it does not lead to
a space volume divergence
like the translation zero mode factor (\ref{2.16d}); 
the parameter describing this symmetry takes values in
$[0,2\pi)$); cf.\ (\ref{6.6})).

The product over the continuum $\omega^{(\perp)}_m=\omega^{(\perp)}(k)$,
$k\in I\!\!R$ in (\ref{5.7d}) is rewritten as an integral over 
$\Phi(k^2;\beta_T,\mu)$ defined in (\ref{vac5}) weighted by the density 
$\rho_{\perp}(k;\kappa^2)$ of the scattering states of eq.\ (\ref{5.9}),
\begin{equation}
\prod_{k\in I\!\!R}\frac{1}{2\sinh\frac{\omega^{(\perp)}(k)\beta_T}{2}}
=\exp\left(-\beta_T\int_{-\infty}^{\infty}dk\,
\rho_{\perp}(k;\kappa^2)\Phi(k^2;\beta_T,\mu)\right)
\label{5.28ca}
\end{equation}

We are now able to write down the contribution of the perpendicular fluctuations
to the transition rate. From (\ref{vac6}) with $i=1$ and 
(\ref{5.7e},\ref{5.28a},\ref{5.28b},\ref{5.28c},\ref{5.28ca}) we have
\begin{eqnarray}
& & \hspace{-16mm}\sqrt{
\det\left(\frac{\hat{{\cal G}}_{\perp}^{(0)}}{\hat{{\cal G}}_{\perp}^{(1)}}
\right)} \nonumber \\
& = &
\frac{{\cal N}_{\perp}}{8i\sin\left(\frac{|\omega_-(\kappa^2)|\beta_T}{2}
\right)\sinh\left(\frac{\omega_-(\kappa^2)\beta_T}{2}\right)}
\nonumber \\
& & \qquad\quad \times
\exp\left(-\beta_T\int_{-\infty}^{\infty}dk\,
\left(\rho_{\perp}(k;\kappa^2)-\rho_0(k)\right)\Phi(k^2;\beta_T,\mu^2)\right)
\label{5.38}
\end{eqnarray}

The integral is performed using the analogue of formula (\ref{2.16g}),
\begin{equation}
\int_{-\infty}^{\infty}dk\,
\left(\rho_{\perp}(k;\kappa^2)-\rho_0(k)\right)\Phi(k^2;\beta_T,\mu) = 
\frac{1}{2\pi}\int_{-\infty}^{\infty}dk\,\Phi(k^2;\beta_T,\mu)\frac{d}{dk}
\delta_{\perp}(k;\kappa^2)
\label{5.38a}
\end{equation}
where $\delta_{\perp}(k;\kappa^2)$ is again the phase shift between incoming
and outgoing plane waves in the scattering states of (\ref{5.9}),
\begin{equation}
\delta_{\perp}(k;\kappa^2)=
\arg\frac{\Gamma(-ik)\Gamma(1-ik)}{\Gamma(-ik-s_{\perp}(k;\kappa^2))
\Gamma(-ik+s_{\perp}(k;\kappa^2)+1)}.
\label{5.38aa}
\end{equation}
Using the decomposition 
$\Phi(k^2;\beta_T,\mu)=\Phi_1(k^2;\mu)+\frac{1}{\beta_T}\Phi_2(k^2,\beta_T\mu)$
(\ref{vac5a}), it is easy to show that the integral
\begin{equation}
\int_{-\infty}^{\infty}dk\,\Phi_1(k^2;\mu)\frac{d}{dk}
\delta_{\perp}(k;\kappa^2)
\label{5.39}
\end{equation}
is again a divergent zero point energy contribution which has to be 
renormalized. With $\Phi_2(k^2,\beta_T\mu)$, we define the function 
($a=\beta_T\mu$)
\begin{equation}
h_{\perp}(a,\kappa^2):=\frac{1}{2\pi}\int_{-\infty}^{\infty}
\ln\left(1-e^{-a\sqrt{1+k^2}}\right)
\frac{d\delta_{\perp}(k;\kappa^2)}{dk}dk
\label{5.40}
\end{equation}
which is discussed in Appendix \ref{app4}. Its high temperature behaviour 
for $\kappa^2>0$ is 
given by (\ref{h20}),
\begin{equation}
h_{\perp}(\beta_T\mu;\kappa^2)\sim \frac{\pi m(\kappa^2)}{6}\frac{1}{\beta_T\mu}
>0
\quad \mbox{for}\; \beta_T\mu\rightarrow 0.
\label{5.41}
\end{equation}
This linear growth with temperature results from the large $k$ behaviour of the
phase shift (\ref{h12}),
\begin{equation}
-\delta_{\perp}(k;\kappa^2) \stackrel{\textstyle<}{\sim}
m(\kappa^2)k+b(\kappa^2) \quad \mbox{for}\;  k \rightarrow \infty.
\label{5.41a}
\end{equation}
In the Skyrme--less limit ($\kappa^2=0\Rightarrow s_{\perp}(k;\kappa^2=0)=2$), 
this phase shift tends towards 
a constant, and the high temperature leading term of 
the continuum fluctuation contributions is again logarithmic (\ref{h7b}),
\begin{equation}
h_{\perp}(a;\kappa^2=0) \sim - 2 \ln(\beta_T\mu)>0 \quad \mbox{for} \quad
\beta_T\mu\longrightarrow 0.
\label{5.41b}
\end{equation}

Finally, the perpendicular fluctuation contribution to the
transition rate (\ref{2.21b}) is
\begin{equation}
\sqrt{
\det\left(\frac{\hat{{\cal G}}_{\perp}^{(0)}}{\hat{{\cal G}}_{\perp}^{(1)}}
\right)}=
\frac{{\cal N}_{\perp}}{8i\sin\left(\frac{|\omega_-(\kappa^2)|\beta_T\mu}{2}
\right)\sinh\left(\frac{\omega_+(\kappa^2)\beta_T\mu}{2}\right)}
e^{-h_{\perp}(\beta_T\mu,\kappa^2)}
\label{5.42}
\end{equation}

\section{The thermal transition rate due to the spha\-le\-ron: 
Skyrme--less limit
and range of validity of the calculation}

\label{sectionrange}

So far we have evaluated explicitly the one--loop fluctuation contributions to 
the perturbative transition
rate (\ref{2.21b}). The classical (zeroth
order) part of this relation,
\begin{equation}
\Gamma = -\frac{|\varrho|}{\pi}e^{-\beta_T\CE[\vec{\varphi}_1]}
\mbox{Im}\sqrt{\det\left(
\frac{(\hat{{\cal G}}_{\parallel}^{(0)}}{(\hat{{\cal G}}_{\parallel}^{(1)}}
\right)
\det\left(\frac{\hat{{\cal G}}_{\perp}^{(0)}}{\hat{{\cal G}}_{\perp}^{(1)}}
\right)}
\label{7.0}
\end{equation}
is given by the exponential of the 
classical action of the sphaleron (\ref{2.12}), i.e.
\begin{equation}
S_{sph}(\beta_T):=S_e[\vec{\varphi}_1] = \beta_T\lambda_1
\int_{-\infty}^{\infty} \left[\frac{1}{2} (f'(x))^2 + 
\mu^2V(\cos f(x)) \right] dx =8\beta_T\mu\lambda_1.
\label{7.1}
\end{equation}

Inserting now (\ref{2.16k},\ref{5.42}) and (\ref{7.1}) into
(\ref{7.0}) (where the damping factor $|\varrho|$ for large 
$\lambda_1$ --- which corresponds to the weak coupling limit of the original
$O(3)$--$\sigma$ model ---
may be approximated by the modulus of the negative eigenvalue,
$|\varrho|=|\omega_-(\kappa^2)|$), the transition rate per unit volume given by the sphaleron is
\begin{equation}
\frac{\Gamma}{L}=\frac{|\omega_-(\kappa^2)|\mu}{\pi}\cdot
\frac{\frac{\lambda_1}{\sqrt{3}\beta_T}
\sqrt{1+\frac58\kappa^2}}{\sin\left(
\frac{|\omega_-(\kappa^2)|\beta_T\mu}{2}
\right)\sinh\left(\frac{\omega_+(\kappa^2)\beta_T\mu}{2}\right)}
\cdot\exp\left\{-8\beta_T\mu\lambda_1-h(\beta_T\mu;\kappa^2)\right\}
\label{7.2}
\end{equation}
with the continuum fluctuation contributions described by the function
\begin{equation}
h(\beta_T\mu;\kappa^2):=h_{\parallel}(\beta_T\mu)+
h_{\perp}(\beta_T\mu;\kappa^2).
\label{7.2a}
\end{equation}
The prefactors of the exponential originate from the transition rate damping
constant (first factor), the zero modes (numerator of the second factor)
and the nonzero discrete modes of the fluctuation spectrum (denominator
of the second factor).

The Skyrme--less limit of the model is given by $\kappa^2\rightarrow 0$. In
this limit, there is no discrete eigenmode with positive eigenvalue
$\omega_+(\kappa^2)$, the contribution 
$\frac{1}{\sinh\left(\frac{\omega_+(\kappa^2)\beta_T\mu}{2}\right)}$
being absorbed in the continuum
fluctuation contribution,
\begin{equation}
\frac{\exp\{-h(\beta_T\mu;\kappa^2)\}}
{2\sinh\left(\frac{\omega_+(\kappa^2)\beta_T\mu}{2}\right)}
\stackrel{\kappa^2\rightarrow 0}{\longrightarrow}
\exp\left\{-h(\beta_T\mu;\kappa^2=0)\right\}.
\label{7.2b}
\end{equation}
With $\left(\omega_-(\kappa^2=0)\right)^2=-3$, the Skyrme--less transition
rate is given by
\begin{equation}
\left.\frac{\Gamma}{L}\right|_{\kappa^2=0} = \frac{2\lambda_1\mu}{\pi\beta_T}
\frac{1}{\sin\left(
\frac{\sqrt{3}\beta_T\mu}{2}\right)}
\exp\left\{-8\beta_T\mu\lambda_1-h(\beta_T\mu;\kappa^2=0)\right\}
\label{7.2c}
\end{equation}
which is exactly the result of Mottola and Wipf (eq.\ (5.10) in their paper),
using our notation ($\lambda_1\leftrightarrow\frac{1}{g^2}$, 
$\omega^2\leftrightarrow\mu^2$).

Next, we discuss the {\em temperature range in which the sphaleron
calculation is valid}. We will see that this range is influenced
by the Skyrme term. Since we performed a
one--loop quantum approximation around the classical sphaleron solution
using the idea that the imaginary part of the
partition function describes transition rates, this range of validity is 
restricted by two technical requirements:
\begin{enumerate}
\item \label{cond1} 
There has to be exactly one negative eigenvalue of the full Gaussian
fluctuation operator. This leads to a lower bound.
\item \label{cond2} 
The one loop fluctuation contributions must not be greater than the 
zeroth approximation given by the classical Boltzmann term. This leads to an 
upper bound.
\end{enumerate}

The negative eigenvalue of the {\em static} fluctuation operator is
(\ref{5.21}), i.e.
\begin{equation}
\omega_-^2(\kappa^2)= \frac{-\left(20\kappa^2+\frac{11}{2}\right)+
\frac12\sqrt{25+112\kappa^2+64(\kappa^2)^2}}{(1+4\kappa^2)^2}<0 
\label{7.2d}
\end{equation}  
and the corresponding eigenvalues of the
full Gaussian fluctuation operator are given by
\begin{equation}
\left(\frac{2\pi n}{\beta_T}\right)^2+
\left(\omega_-(\kappa^2)\right)^2\mu^2.
\label{7.2e}
\end{equation}
$n=0$ always yields a negative eigenvalue, the eigenvalues for $n>0$ thus
have to be nonnegative. Inserting $n=1$, one obtains from condition \ref{1}
\begin{equation}
k_B T > \frac{\left|\omega_-(\kappa^2)\right|\mu}{2\pi} =:
k_B T^{(-)}(\mu,\kappa^2).
\label{7.2f}
\end{equation}
Without Skyrme term, this lower bound is $T^{(-)}(\mu,\kappa^2=0)=
\frac{\sqrt{3}}{2\pi}\mu$. With Skyrme term, it
decreases with increasing Skyrme coupling
(since $\left|\omega_-(\kappa^2)\right|$ decreases with increasing
$\kappa$). But for these lower temperatures, the exponential suppression 
due to the Boltzmann factor increases.

To analyse the second condition, one has to compare the classical action of 
the sphaleron, $S_e[\vec{\varphi}_1]=8\beta_T\mu\lambda_1$ with the
quantum effects of the linear fluctuations around it, given by
the continuum fluctuation function $h$ in the exponent of (\ref{7.2}) and the
zero and discrete eigenmode factors in that formula. In fact, it is sufficient
to compare $S_e[\vec{\varphi}_1]$ with $h(\beta_T\mu;\kappa^2)$, because
for high temperatures (small $\beta_T$), the discrete mode prefactors
of (\ref{7.2}) yield only logarithmic $\beta_T\mu$ contributions in the
exponent, and $h$ itself has at least a logarithmic 
$\beta_T\mu\rightarrow 0$ behaviour. The upper temperature bound is thus
given by the condition
\begin{equation}
8\beta_T\mu\lambda_1 \gg h(\beta_T\mu;\kappa^2)
\label{7.2g}
\end{equation}
in which we ignore all numerical prefactors to obtain a rough estimate
of the upper temperature bound $T^{(+)}(\mu,\lambda_1;\kappa^2)$.

In the Skyrme--less case $\kappa^2=0$, the high temperature behaviour
of $h$ is given by (\ref{2.16j}), (\ref{5.41b}), i.e.\ 
$h(\beta_T\mu,\kappa^2=0)\sim -3\ln(\beta_T\mu)$, $\beta_T\mu\rightarrow 0$.
For small arguments, any negative power function grows faster than the
negative logarithm, $\lim_{a\rightarrow 0}\frac{(-\ln(a))}{a^{\gamma}}=0$ 
for all $\gamma>0$. This means that for any $\gamma>0$, 
there must exist a number $a^*(\gamma)$ such that
for all $a<a^*(\gamma)$, we have $-\ln(a)<a^{-\gamma}$. Inserting this 
into (\ref{7.2g}) without numerical factors yields
\begin{equation}
k_B T \ll \mu\lambda_1^{\frac{1}{\gamma+1}}
\label{7.2h}
\end{equation}
The upper temperature limit therefore increases with $\lambda_1$: High
temperatures are allowed for large $\lambda_1$ which is the small coupling
limit of the original $O(3)$--$\sigma$ model. This means we consider small
$\beta_T\mu=a$ in the approximation of the logarithm $-\ln(a)<a^{-\gamma}$,
therefore small $\gamma\neq0$ are allowed in this approximation.
Thus, the upper temperature bound in the ``small coupling
limit'' $\lambda_1\rightarrow\infty$ is given by
\begin{equation}
k_B T \ll \lambda_1\mu =: k_B T^{(+)}(\lambda_1\mu;\kappa^2=0).
\label{7.2hh}
\end{equation}

Next, including the Skyrme term changes the high temperature behaviour of
$h$ drastically. In view of (\ref{5.41}) we have
\begin{equation}
h(\beta_T\mu;\kappa^2>0)\sim \frac{\pi m(\kappa^2)}{6}\frac{1}{\beta_T\mu}
\quad \mbox{for}\; \beta_T\mu\rightarrow 0.
\label{7.2i}
\end{equation}
Inserting this into (\ref{7.2g}) and again ignoring numerical prefactors
immediateley yields
\begin{equation}
k_B T \ll \mu \sqrt{\frac{\lambda_1}{m(\kappa^2)}} =: 
k_B T^{(+)}(\lambda_1,\mu;\kappa^2>0)
\label{7.2j}
\end{equation}
For fixed $\kappa^2>0$, this upper bound is considerably smaller than
the Skyrme--less upper bound (\ref{7.2h}) in the ``small coupling limit''.
One main effect of the Skyrme term in the sphaleron calculation
of the thermal transition rate is thus to reduce the range of validity of the
calculation. In the Skyrme--less limit, 
$T^{(+)}(\lambda_1,\mu;\kappa^2>0)$
in eq.\ (\ref{7.2i}) increases as $m(\kappa^2)$ decreases for
$\kappa^2\rightarrow 0$ which leads back to the Skyrme--less upper
temperature bound condition (\ref{7.2h}).

\section{Discussion and conclusions}

\label{sectionconc}

After the detailed discussion of the sphaleron and the fluctuations 
around it in the Skyrmed reduced $O(3)$--$\sigma$ model we enquire about
the physical implications of the results. 

First of all, from a very naive point of view, one might conclude 
that eq.\ (\ref{7.2}) describes thermal transitions over the barrier 
separating the degenerate vacua which are exponentially suppressed at high
temperatures. This would destroy the most important property of the
sphaleron induced transition --- namely that of not leading to
exponential suppression ---
compared to that induced by the (vacuum) instanton which is always exponentially
suppressed. But eq.\ (\ref{7.2}) does not demand this interpretation, because
in the temperature region where this exponential suppression becomes
dominant, this formula is no longer valid for the reasons discussed
in the last section.

At very high temperatures, $T>T^{(+)}$, the sphaleron calculation thus breaks 
down. In realistic models with spontaneous symmetry breaking (and particularly
in the electroweak model), this is cured by the restoration of symmetry 
above $T_{rest}\approx T^{(+)}$ (which is indicated in the diagrams of
Fig.\ \ref{figregions1} and Fig.\ \ref{figregions2}). At
these temperatures, the fluctuations become very large and the system becomes 
subject to
disorder. It is clear that in this temperature range the transition rate
becomes unsuppressed and there is no need to worry about the sphaleron which
ceases to exist because it needs the broken symmetry.
In $O(3)$--$\sigma$ models (with or without Skyrme term), where there is no 
Higgs--mechanism of spontaneous symmetry breaking, one has to add a 
symmetry breaking potential $\mu V(\phi^3)$ by hand.  One can show that the 
explicitly broken symmetry of the $O(3)$--$\sigma$ model is {\em not} 
restored at high temperatures, so that for very high temperatures $T>T^{(+)}$,
both the Skyrme--less and the Skyrmed reduced $O(3)$--$\sigma$ model cease to
be physically meaningful theories. The only difference in the very high 
temperature behaviour is therefore that $T^{(+)}$ decreases with increasing 
Skyrme coupling $\kappa^2$, but this is what one expects on physical grounds, 
because
at high temperatures, the Skyrme term (having the highest number of derivatives
in the Lagrangian) dominates the theory with large fluctuation effects which
would lead to symmetry restoration also at lower temperatures in a realistic
theory.

Finally, we consider the situation in the lower temperature region around
$T^{(-)}$. As already remarked in the introduction, the Skyrme--less limit of
the model (i.e.\ the original reduced $O(3)$--$\sigma$ model) is a type II 
theory with dominant transition processes as indicated in Fig.\ 
\ref{figregions2}. This changes completely for $\kappa^2>0$ 
since the Skyrme term
enables the existence of localised instantons with Euclidean energy (\ref{1b})
\cite{Kuznetsov}
\begin{equation}
2S_0 = 8\pi\lambda_1 w(\kappa), \qquad w(\kappa)=1+\CO(\kappa)
\label{8.1}
\end{equation}
(the function $w(\kappa)$ has to be evaluated numerically since the 
vacuum instanton is
not known explicitly \cite{modelidea}) and corresponding periodic instantons
with $\frac{\partial\CE_{inst}}{\partial \beta_T}<0$.In the limit 
$\kappa^2\rightarrow 0$, there is thus a discontinuous change in the {\em type}
of instantons involved 
(whereas the {\em action} of the vacuum instanton is continuous
for $\kappa^2\rightarrow 0$, the action of the constrained instanton 
in the Skyrme--less theory can be estimated to be $2S_0 = 8\pi\lambda_1$
\cite{Habib}). In the case of the sphaleron of the Skyrmed model, there
is a smooth crossover to the Skyrme--less situation. The sphaleron is 
only quantitatively changed 
by the Skyrme term, whereas the instanton changes 
qualitatively. In particular, the Euclidean action $S_{sph}^{(-)}$ of the
sphaleron at the lower temperature limit $T^{(-)}$,
\begin{equation}
S_{sph}^{(-)}=\frac{16\pi\lambda_1}{|\omega_-(\kappa^2)|}
\label{8.2}
\end{equation}
depends smoothly on the Skyrme coupling $\kappa^2$. This means that the type II
condition 
\begin{equation}
2S_0<S_{sph}^{(-)}\quad \Leftrightarrow \quad T_0>T^{(-)}
\label{8.3}
\end{equation}
which is true for $\kappa^2=0$ remains true for $\kappa^2>0$ due to the
smooth $\kappa^2$--dependence of both actions, whereas the
type of instantons changes discontinuously
from those in type II theories to those in type I theories. The Skyrmed reduced
$O(3)$--$\sigma$ model is therefore neither of type I nor of type II at least
for a weak Skyrme coupling \cite{Kuznetsov}. 
Whether it may become a type I theory with increasing
Skyrme coupling depends on the function $w(\kappa)$ which would have to satisfy
the condition $w(\kappa)|\omega_-(\kappa^2)|>2$ for some $\kappa^2$. 
But this is very unlikely to happen, because as shown in
Fig.\ \ref{fig1}, $|\omega_-(\kappa^2)|$
decreases rapidly to $0$ as $\kappa^2$ increases.

\section*{Acknowledgements}

This work was supported in part by a Forbairt(Ireland) grant, project No.
IC/97/019.

The authors are indebted to A. Wipf (Jena) for detailed discussions and to
V.A. Rubakov (Moscow) also for a critical reading of the manuscript with
valuable comments.

\section*{Appendix}
\begin{appendix}

\section{Zero modes of the $O(3)$ model}

\label{section6}
\label{app2}

The spectra of the operators $\CH_{\parallel}$, $\CH_{\perp}$
each contain one zero eigenvalue which appears in the evaluation of the
fluctuation integrals (\ref{2.19},\ref{2.20}) and yields  divergences in 
(\ref{2.16c},\ref{5.7d}). In this Appendix, we discuss the origin of these 
``zero modes'' \cite{zero}.

For this reason we analyse the symmetries of the model (\ref{1})
which are broken by the classical solution 
$\vec{\varphi}_1(x)=\left(\sin f(x),0,\cos f(x)\right)$. 
We consider the following symmetries of the model (\ref{1}):
\begin{enumerate}
\item space translations:
\begin{eqnarray}
{\bf T}[\vec{\phi},a](\tau,x) & = & \vec{\phi}(\tau,x+a) \nonumber \\
& = & \vec{\phi}(\tau,x)
+\frac{\partial \vec{\phi}}{\partial x}(\tau,x)\cdot a +\CO(a^2)
\quad \mbox{for} \quad a\rightarrow 0
\label{6.1}
\end{eqnarray}
(There is also a time translation symmetry, but this is 
not broken by a static solution.)
\item internal $SO(2)$--rotations:
\begin{eqnarray}
{\bf R}[\vec{\phi},\alpha](\tau,x) & = & \left(\begin{array}{ccc}
\cos \alpha & -\sin \alpha & 0 \\ \sin \alpha & \cos \alpha & 0
\\ 0 & 0 & 1 \end{array} \right)
\vec{\phi}(\tau,x)\nonumber \\
& = & \vec{\phi}(\tau,x) + \left(\begin{array}{ccc}
0 & -1 & 0 \\ 1 & 0 & 0 \\ 0 & 0 & 1 \end{array} \right) 
\vec{\phi}(\tau,x)\cdot \alpha +\CO(\alpha^2)
\quad \mbox{for} \quad \alpha\rightarrow 0 \nonumber \\
\label{6.2}
\end{eqnarray}
\end{enumerate}
The classical solution $\vec{\varphi}_1(x)=(\sin f(x), 0, \cos f(x))$ breaks
this invariance because 
\begin{equation}
{\bf T}[\vec{\varphi}_1,a](\tau,x)\equiv\vec{\varphi}_1(x+a)
\neq \vec{\varphi}_1(x), \qquad {\bf R}[\vec{\varphi}_1,\alpha](\tau,x)
\neq \vec{\varphi}_1(x). 
\label{6.3}
\end{equation}
The moduli space \cite{Moduli} of the parameters $a$, $\alpha$ describing 
finite energy solutions is therefore
topologically ${\cal M}=I\!\!R^1\times{\bf\rm S}^1$ and
one can show that $\frac{\partial {\bf T}}
{\partial a}[\vec{\varphi}_1,0](x)$ and $\frac{\partial {\bf R}}
{\partial \alpha}[\vec{\varphi}_1,0](x)$ are zero modes of the 
static Gaussian 
fluctuation operator $\left[\frac{\delta^2\CE[\vec{\phi}]}
{\delta\phi^i(x)\delta\phi^j(x')}\right]_{\vec{\varphi}}$ by expanding 
\begin{equation}
\CE\left[\vec{\varphi}_1+
\frac{\partial {\bf T}}{\partial a}[\vec{\varphi}_1,0] \right] \qquad \mbox{and}
\quad
\CE\left[\vec{\varphi}_1+
\frac{\partial {\bf R}}{\partial a}[\vec{\varphi}_1,0] \right]
\label{6.4}
\end{equation}
around $\vec{\varphi}_1$. The problem here is that one has to be very careful
with the constraint $\vec{\phi}^2=1$, and it is better to consider the
broken symmetries in the parametrization $\vec{\varphi}(\vec{f})$. Therefore
we first express the symmetries (\ref{6.1}), (\ref{6.2}) in terms of 
the parameter fields. This is done by expanding the ansatz (\ref{1.4a})
in small quantities $\delta \vec{f}$ around 
$\vec{f}=\left(f^{(\parallel)},f^{(\perp)}\right)$ 
and then comparing the result
with the infinitesimal form of the symmetries given by (\ref{6.1}),
(\ref{6.2}), also written in terms of the parameter fields. The result
is
\begin{enumerate}
\item for space translations:
\begin{equation}
{\bf T}_f[\vec{f},a](\tau,x) 
= \vec{f}(\tau,x)
+\frac{\partial \vec{f}}{\partial x}(\tau,x)\cdot a +\CO(a^2)
\quad \mbox{for} \quad a\rightarrow 0
\label{6.5}
\end{equation}
\item for internal $SO(2)$--rotations:
\begin{eqnarray}
{\bf R}_f[\vec{f},\alpha](\tau,x)
= \vec{f}(\tau,x) + \left(\begin{array}{c}
-f^{(\perp)}(\tau,x)\cos f^{(\parallel)}(\tau,x) \\ \left(1+(f^{(\perp)}(\tau,x))^2\right)\sin f^{(\parallel)}(\tau,x)
\end{array} \right) 
\cdot \alpha +\CO(\alpha^2) \nonumber \\
\quad \mbox{for} \quad \alpha\rightarrow 0 \quad
\label{6.6}
\end{eqnarray}
\end{enumerate}
In terms of the parametrization $\vec{f}$, the classical solution 
$\vec{\varphi}$ is given by $\vec{f}(x)=\vec{f}_0(x)=(f(x),0)$. This yields
the following two zero modes of the Gaussian fluctuation operator in 
$\vec{f}$--parametrization,
\begin{eqnarray}
& & \frac{\partial {\bf T}_f}
{\partial a}[\vec{f}_0,0] = \left(\begin{array}{c}
f' \\ 0 \end{array}\right), \qquad f'(x)=2\mu\sech(\mu x) \label{6.7a} \\
& & \frac{\partial {\bf R}_f}
{\partial \alpha}[\vec{f}_0,0]
 = \left(\begin{array}{c}
0 \\ \sin f \end{array}\right), \qquad \sin f(x) = -2\sech(\mu x)\tanh(\mu x)
\label{6.7b}
\end{eqnarray}
Obviously, they are proportional to the zero eigenvalue
eigenfunctions of the fluctuation equations (\ref{2.16b},\ref{5.3}), normalised
with respect to the appropriate scalar products (\ref{2.16ca},\ref{5.5}),
\begin{eqnarray}
& & \psi^{(\parallel)}_0(x) = \sqrt{\frac{\mu}{2}}\sech(\mu x), \qquad
\left\langle\psi^{(\parallel)}_0,\psi^{(\parallel)}_0\right\rangle_{\parallel}
=1 \label{6.7aa} \\
& & \psi^{(\perp)}_1(x) = \sqrt{\frac{15\mu}{10+16\kappa^2}}
\sech(\mu x)\tanh(\mu x), \qquad
\left\langle\psi^{(\perp)}_1,\psi^{(\perp)}_1\right\rangle_{\perp}=1
\label{6.7bb}
\end{eqnarray}

Next, we need a convenient way to take these zero modes into account in the 
Fou\-ri\-er--Mat\-su\-ba\-ra evaluation of the fluctuation integrals 
(\ref{2.16c},\ref{5.7d}). Usually, on uses the method of collective coordinates
which means that one treats the parameters which describe the breakdown
of a global symmetry of the theory in a static solution ($a$ and $\alpha$ 
in the case of the $O(3)$ model) as additional dynamical degrees of freedom,
i.e.\ one considers
\begin{equation}
\vec{\phi}(\tau,x)=
{\bf R}[{\bf T}[\vec{\varphi},a(\tau)],\alpha(\tau)](\tau,x)
\label{zero1}
\end{equation}
and thus adds two redundant
degrees of freedom $a$, $\alpha$. 
One can show that the resulting theory is a gauge 
theory, i.\ e.\ the Euclidean action of the theory in terms of the new 
``fields'' $S_e[\vec{v},a,\alpha]$ ($\vec{v}$ denoting again the
parameter fields of the fluctuations) is invariant under some (time--)local
transformations of $\vec{v}$, $a$, $\alpha$. Therefore, the usual techniques
for path--integral quantization of gauge theories (Faddeev--Popov \cite{FadPop}
or BRST \cite{brst}) can
be applied to evaluate the partition function ${\cal Z}_1$ around
$\vec{\varphi}_1$. A convenient gauge--fixing condition is to exclude 
fluctuations parallel to the zero modes.
This procedure is well--known and discussed in detail in the literature, even 
in the particular context of $O(3)$ models, using classical Faddeev--Popov
\cite{FadPopSol} or more modern BRST and BV methods \cite{brstSol,bvSol}. 
Therefore, we do not discuss the details. 

Here, we use a simpler semiclassical method \cite{MottWipf} to compute the
zero mode contributions to the partition function, i.e.\ to treat them in
fluctuation integrals (\ref{2.16c},\ref{5.7d}). In the harmonic oscillator
decomposition of the Gaussian fluctuation operators 
${\cal G}_{\parallel}^{(1)}$, 
${\cal G}_{\perp}^{(1)}$, 
each zero mode contributes one harmonic oscillator mode
with zero frequency, i.\ e.\ a free mode. The classical contribution of
a free mode to the partition function is
\begin{equation}
\int \frac{dq dp}{2\pi}e^{-\beta_T\frac{p^2}{2}} = \frac{1}{\sqrt{2\pi\beta_T}}
\int dq.
\label{zero3}
\end{equation}
The ``coordinate'' $q$ of this free mode has to be identified with the
zero mode coefficients $c^{(\parallel)}_0$, $c^{(\perp)}_1$ in the
harmonic oscillator decompositions of the parallel and perpendicular
fluctuations, (\ref{2.16caq}) and (\ref{5.6}), respectively. Therefore, 
we consider the fluctuations in direction of the zero modes. Let
${\bf pr}_0$ be the projector from the space of fluctuations
$\vec{v}(\tau,x)$ on the subspace of zero mode fluctuations.
Since the zero modes are those which result from fluctuations in the
collective coordinates, expanding the parameter field version of (\ref{zero1})
\begin{equation}
\vec{f}(\tau,x)={\bf R}_f[{\bf T}_f[\vec{f}_0,a(\tau)],\alpha(\tau)](\tau,x) 
\label{zero4}
\end{equation}
around the classical static solution 
$\vec{f}_0={\bf R}_f[{\bf T}_f[\vec{f}_0,0],0]$ (small $a$, $\alpha$) yields
\begin{eqnarray}
\vec{f}(\tau,x) & = &
{\bf R}_f[{\bf T}_f[\vec{f}_0,0],0](\tau,x) \nonumber \\
& & {} +
{\bf R}_f\left[\frac{d{\bf T}_f}{da}[\vec{f}_0,0],0\right](\tau,x)\cdot
a(\tau)+
\frac{d{\bf R}_f}{d\alpha}[{\bf T}_f[\vec{f}_0,0],0](\tau,x)\cdot\alpha(\tau) 
\nonumber \\
& = & \vec{f}_0(x)+ \frac{d{\bf T}_f}{da}[\vec{f}_0,0](\tau,x)\cdot a(\tau) +
\frac{d{\bf R}_f}{d\alpha}[\vec{f}_0,0](\tau,x)\cdot\alpha(\tau)   \nonumber \\
& = & \vec{f}_0(x) + \sqrt{\lambda_1}\,{\bf pr}_0\{\vec{v}(\tau,x)\}
\label{zero5}
\end{eqnarray}
with
\begin{equation}
{\bf pr}_0\{\vec{v}(\tau,x)\}=\frac{d{\bf T}_f}{da}[\vec{f}_0,0](\tau,x)
\cdot a(\tau) +
\frac{d{\bf R}_f}{d\alpha}[\vec{f}_0,0](\tau,x)\cdot\alpha(\tau).
\label{zero6}
\end{equation}
(The factor $\lambda_1$ appears because an overall factor $\lambda_1$ is 
extracted from the fluctuation expansion (\ref{2.9}--\ref{2.11})).

But from the decompositions (\ref{2.16caq}) and (\ref{5.6}), the parallel
and perpendicular components of the projection operator ${\bf pr}_0$
are given by 
${\bf pr}^{(\parallel)}_0=\left\langle\cdot,\psi^{(\parallel)}_0
\right\rangle_{\parallel}\psi^{(\parallel)}_0$ and 
${\bf pr}^{(\perp)}_0=\left\langle\cdot,\psi^{(\perp)}_1
\right\rangle_{\perp}\psi^{(\perp)}_1$, so that with (\ref{6.7a},\ref{6.7b}):
\begin{equation}
\sqrt{\lambda_1}\,{\bf pr}_0\{\vec{v}(\tau,x)\}=\sqrt{\lambda_1}\,\left(
\begin{array}{c} c^{(\parallel)}_0(\tau)\psi^{(\parallel)}_0(x) \\
c^{(\perp)}_1(\tau)\psi^{(\perp)}_1(x)\end{array}\right) = 
\left(
\begin{array}{c} a(\tau) f'(x) \\ \alpha(\tau) \sin f(x) \end{array}\right).
\label{zero7}
\end{equation}
Thus we have
\begin{eqnarray}
\int dq & \longrightarrow & \sqrt{\lambda_1}\,\int dc^{(\parallel)}_0 = 
\sqrt{\lambda_1}\,\Delta c^{(\parallel)}_0
=\sqrt{\lambda_1}\,
\sqrt{\langle f',f' \rangle_{\parallel}}\Delta a \label{zero8a} \\
\int dq & \longrightarrow & \sqrt{\lambda_1}\,\int dc^{(\perp)}_1 = 
\sqrt{\lambda_1}\,\Delta c^{(\perp)}_1
=\sqrt{\lambda_1}\,
\sqrt{\langle \sin f, \sin f \rangle_{\perp}} \Delta \alpha \label{zero8b}
\end{eqnarray}
with 
\begin{equation}
\langle f',f' \rangle_{\parallel} = 4\mu^2, \qquad
\langle \sin f, \sin f \rangle_{\perp} =\frac{10+16\kappa^2}{15}\frac{4}{\mu}
\label{zero9}
\end{equation}
$\Delta a$, $\Delta \alpha$ are the values the parameters describing
the broken symmetry can take: For rotations, it is clear that 
$\Delta \alpha =2\pi$. To avoid a volume divergence due to the translation 
parameter $a\in I\!\!R$, we restrict the model to a finite space volume
such that $\Delta a = L$. Therefore the zero mode factors are
\begin{equation}
L{\cal N}_{\parallel}=2L\sqrt{\lambda_1}\sqrt{\frac{\mu}{\pi\beta_T}}, \quad
{\cal N}_{\perp}=4\sqrt{\lambda_1}
\sqrt{\frac{\pi}{3\beta_T\mu}}\sqrt{1+\frac85\kappa^2}.
\label{zero10}
\end{equation}

\section{The Schr\"odinger equation with $\mbox{sech}^2$--potential}

\label{section3}
\label{app3}

All attempts to evaluate the fluctuation factors (\ref{2.19},\ref{2.20}) 
lead to the eigenvalue differential equation
\begin{equation}
\psi'' + (E+U\sech^2y)\psi=0
\label{3.1}
\end{equation}
with $\psi=\psi(y)$ which we discuss in this section \cite{Schrodinger}. 
There are three different cases to be distinguished:
\begin{enumerate}
\item $U>0$: then it is possible to have
\begin{enumerate}
\item $E<0$, i.\ e.\ bound states (discrete spectrum) \label{U>0-E<0}
\item $E>0$, i.\ e.\ scattering states (continuous spectrum) \label{U>0-E>0}
\end{enumerate}
\item $U<0$, then $E>0$ and only scattering states (continuous spectrum)
exist. \label{U<0}
\end{enumerate}

We first set $\xi=\tanh y$ to get (now 
$\psi=\psi(\xi)$):
\begin{equation}
\frac{d}{d\xi}\left[(1-\xi^2)\frac{d\psi}{d\xi}\right]
+\left(U+\frac{E}{1-\xi^2}\right)\psi
=0
\label{3.2}
\end{equation}

Assuming $U>0$, we  
first compute the discrete spectrum for which $E<0$ is valid.
We may thus define
\begin{equation}
\epsilon := \sqrt{-E}, \qquad \epsilon > 0
\label{3.3}
\end{equation}
and 
\begin{equation}
U=s(s+1), \qquad s=-\frac12 + \sqrt{\frac14+U} > 0
\label{3.4}
\end{equation} 
to obtain from (\ref{3.2}) the differential equation of generalized
Legendre functions,
\begin{equation}
\frac{d}{d\xi}\left[(1-\xi^2)\frac{d\psi}{d\xi}\right]
+\left[s(s+1)-\frac{\epsilon^2}{1-\xi^2}\right]\psi =0
\label{3.5}
\end{equation}
Next we set $\psi(\xi)=(1-\xi^2)^{\frac{\epsilon}{2}}w(\xi)$ which yields
\begin{equation}
(1-\xi^2)\frac{d^2w}{d\xi^2}-2(\epsilon+1)\frac{dw}{d\xi}+
[s(s+1)-\epsilon-\epsilon^2]w=0
\label{3.6}
\end{equation}
Finally, changing the argument to $\chi=\frac12(1-\xi)$, (\ref{3.6}) transforms
to
\begin{equation}
\chi(\chi-1)\frac{d^2w}{d\chi^2}+[(\epsilon+1)-(2\epsilon+2)\chi]
\frac{dw}{d\chi}
-(\epsilon-s)(\epsilon+s+1)w=0
\label{3.7}
\end{equation}
which is a hypergeometric differential equation.

The general form of the hypergeometric differential equation is
\begin{equation}
z(z-1)\frac{d^2\phi}{dz^2}+[(c-(a+b+1)z]\frac{d\phi}{dz}-ab\phi=0
\label{3.8}
\end{equation}
Solutions are given by the {\em hypergeometric functions}
\begin{equation}
F(a,b,c;z)=\sum_{k=0}^{\infty} \frac{(a)_k(b)_k}{(c)_k}\frac{z^k}{k!}
\label{3.9}
\end{equation}
with
\begin{equation}
(a)_k=\frac{\Gamma(a+k)}{\Gamma(a)}=a(a+1)\cdot\ldots\cdot(a+k-1),\qquad 
(a)_0=1.
\label{3.10}
\end{equation}
The series (\ref{3.9}) converges for $|z|<1$,
$z\in C\!\!\!\!I$. It also converges for $z=1$ if $c-a-b>0$.

Comparing (\ref{3.7}) with the general form (\ref{3.8}), we identify
$a=\epsilon-s$, $b=\epsilon+s+1$ and $c=\epsilon+1$, so the solution
to eq.\ (\ref{3.7}) is given by
\begin{equation}
w(\chi)=F(\epsilon-s,\epsilon+s+1,\epsilon+1;\chi).
\label{3.11}
\end{equation}
Inserting all the substitutions, the solution to the original equation
(\ref{3.1}) may be written as
\begin{equation}
\psi(y)=\left(1-\tanh^2y\right)^{\frac{\epsilon}{2}}
F\left(\epsilon-s,\epsilon+s+1,\epsilon+1;\frac12(1-\tanh y)\right).
\label{3.12}
\end{equation}

We want to
compute bound states (discrete eigenvalues for $E<0$). Hence the solutions
(\ref{3.11}) have to be square integrable which means that at least (as all
functions are continuous) 
$\psi(y)\stackrel{|y|\rightarrow\infty}{\longrightarrow}0$. In terms of
the $\chi$--variable, which
is related to the original variable 
$y$ by $\chi=\frac12(1-\tanh y)$, this means that
$w(\chi)$ has to be bounded as  
$\chi\stackrel{y\rightarrow\infty}{\longrightarrow}0$ and
$\stackrel{y\rightarrow-\infty}{\longrightarrow}1$. 
We always have $F(a,b,c;0)=1$, so the limit $y\rightarrow\infty$ is no 
problem. But since $c-a-b=-\epsilon<0$, $F(a,b,c;1)$ is not bounded for general
values of $a,b,c$. We thus have to force 
$F(\epsilon-s,\epsilon+s+1,\epsilon+1;1)<\infty$ which is achieved 
by making the
infinite sum (\ref{3.9}) finite: $(m)_k=0$ for $-m\in I\!\!N_0$ and $k>m$.
This yields the {\em quantization condition} $\epsilon-s=-m\in I\!\!N_0$: $m$
labels the eigenvalues of (\ref{3.5}), $\epsilon_m=s-m$. There are no
eigenvalues for any $m\in I\!\!N_0$ in view of the condition
(\ref{3.3}) $\epsilon>0$ which means $m<s$. The discrete spectrum of eq.\
(\ref{3.1}) (case \ref{U>0-E<0}) is therefore given by
\begin{equation}
\left\{E_m=-\left(\left.\sqrt{\frac14+U}-\frac12-m\right)^2\right|
m\in I\!\!N_0 \quad\mbox{and}\quad T_U(m)>0\right\}
\label{3.13}
\end{equation}
with 
\begin{equation}
T_U(m):=\sqrt{\frac14+U}-\frac12-m,
\label{3.14}
\end{equation}
and the corresponding eigenfunctions are (with $1-\tanh^2y=\mbox{sech}^2y$)
\begin{equation}
\psi_m(y)=\mbox{sech}^{(s-m)}(y)F\left(-m,2s-m+1,s-m+1
;\frac12(1-\tanh y)\right).
\label{3.14a}
\end{equation}

Next, we assume $U<0$. Then, there is only a positive, continuous spectrum
(case \ref{U<0}) with $E>0$, and we substitute
\begin{eqnarray}
k:=\pm\sqrt{E} & \Longrightarrow & (\pm ik)^2=-E
\label{3.15}\\
U=s(s+1), & & s=-\frac12+\sqrt{\frac14+U} \in C\!\!\!\!I.
\label{3.16}
\end{eqnarray}
Performing the same steps as above, the scattering states of eq.\ (\ref{3.1})
for $U<0$ are given by
\begin{equation}
\psi_{\pm,k}(y)=\left(1-\tanh^2y\right)^{\pm\frac{ik}{2}}
F\left(\pm ik-s,\pm ik+s+1,\pm ik+1;\frac12(1-\tanh y)\right)
\label{3.17}
\end{equation}
and the continuous spectrum is
\begin{equation}
\left\{E=k^2|k\in I\!\!R \right\}.
\label{3.18}
\end{equation}
It is obvious that these calculations are also true for $U>0$, $E>0$ (then,
$s$ is real), so that (\ref{3.17}) also describes the scattering states
of the $U>0$ potential which has bound and scattering eigenstates
(case \ref{U>0-E>0}).

Next, we analyse the asymptotic behaviour of the scattering states
(\ref{3.17}). It will be sufficient to consider the scattering state
$\psi_{-,k}(y)$. The prefactor in (\ref{3.17}) behaves like
\begin{equation}
\left(1-\tanh^2 y\right)^{-\frac{ik}{2}} \sim 4^{\frac{ik}{2}} 
e^{\pm ik} \quad \mbox{for} \quad x \longrightarrow \pm \infty
\label{3.19}
\end{equation}
(we ignore the common factor $4^{\frac{ik}{2}}$ in the following.)
To deal with the hypergeometric function, we use (\ref{3.9}), i.\ e.\
$F(a,b,c;z)=1+\CO(z)$ for $z\rightarrow 0$. This works well for
$y\rightarrow +\infty$, because then $\frac12(1-\tanh y)\rightarrow 0$. For
$y\rightarrow -\infty$, we use the following identity:
\begin{eqnarray}
F(a,b,c;z) 
& = & \frac{\Gamma(c)\Gamma(c-a-b)}{\Gamma(c-a)\Gamma(c-b)}
F(a,b,a+b+1-c;1-z) \nonumber \\
& & {}+ \frac{\Gamma(c)\Gamma(a+b-c)}{\Gamma(a)\Gamma(b)}(1-z)^{c-a-b}
F(c-a,c-b,c+1-a-b;1-z) \nonumber \\
\label{3.20}
\end{eqnarray}
which yields, applied to (\ref{3.17})
\begin{eqnarray}
& & \hspace{-1.5cm}
F\left(-ik-s,-ik+s+1,-ik+1;\frac12(1-\tanh y)\right) \nonumber \\
& = & \frac{\Gamma(-ik+1)\Gamma(ik)}{\Gamma(-s)\Gamma(1+s)}
F\left(-ik-s,-ik+s+1,-ik+1;\frac12(1+\tanh y)\right) \nonumber \\
& & {} +\frac{\Gamma(-ik+1)\Gamma(-ik)}{\Gamma(-ik-s)\Gamma(-ik+s+1)}
\left(\frac12(1+\tanh y)\right)^{ik} \nonumber \\
& & \qquad \times F\left(s+1,-s,ik+1;\frac12(1+\tanh y)\right). \label{3.21}
\end{eqnarray}
Inserting
\begin{equation}
\frac12(1\pm\tanh y) \sim e^{\pm 2y} 
\quad \mbox{for} \quad y \longrightarrow \mp \infty
\label{3.22}
\end{equation}
and putting everything together, we get the following asymptotic
behaviour of the scattering states:
\begin{eqnarray}
\psi_{-,k}(y) & \stackrel{y\rightarrow -\infty}{\sim} &
A(k)e^{-iky} + B(k)e^{iky}
\label{3.23} \\
\psi_{-,k}(y) & \stackrel{y\rightarrow +\infty}{\sim} & e^{iky}.
\label{3.24}
\end{eqnarray}
with
\begin{equation}
A(k)=\frac{\Gamma(ik)\Gamma(1-ik)}{\Gamma(-s)\Gamma(1+s)}  
\qquad B(k)=\frac{\Gamma(-ik)\Gamma(1-ik)}{\Gamma(-ik-s)\Gamma(-ik+s+1)}.
\label{3.25}
\end{equation}
Since $k^2>0$, the poles of $B(k)$ are excluded. $A(k)$ vanishes for 
$s\in I\!\!N$. This is the well-known case of the reflection free 
Rosen--Morse potential.

Eqs. (\ref{3.23}, \ref{3.24}) show
that $\psi_{-,k}$ describes incoming particles from the left; 
there are no reflected particles to the right (one can check that the second
independent solution
$\psi_{+,k}$ describes incoming particles from the right). The incoming flux is
normalised to unity. Thus, all important information about the scattering 
solutions resides in the phase shift $\delta$ defined by the transmission
coefficient $B(k)\equiv e^{i\delta}$:
\begin{equation}
\delta=\delta_s(k)=
\arg\frac{\Gamma(-ik)\Gamma(1-ik)}{\Gamma(-ik-s)\Gamma(-ik+s+1)}.
\label{3.26}
\end{equation}
For $s\in I\!\!N$, there is a simple formula for $\delta_s(k)$ \cite{Tsitsi}:
\begin{equation}
\delta_s(k)=-2\sum_{n=1}^s \arctan\left(\frac{k}{n}\right).
\label{3.27}
\end{equation}

\section{(High) temperature behaviour of the functions $h_{\parallel}$ and
$h_{\perp}$}

\label{app4}

In this Appendix, we analyse the small $a=\beta_T\mu$ behaviour of the functions
$h_{\parallel}$ (\ref{2.16i}), $h_{\perp}$ (\ref{5.40}) given by the integrals
\begin{equation}
\frac{1}{2\pi}\int_{-\infty}^{\infty}
\ln\left(1-e^{-a\sqrt{1+k^2}}\right)
\frac{d\delta_{s_{\parallel,\perp}}(k)}{dk}dk
\label{h1}
\end{equation}
with (\ref{3.26})
\begin{equation}
\delta_s(k):=
\arg\frac{\Gamma(-ik)\Gamma(1-ik)}{\Gamma(-ik-s)\Gamma(-ik+s+1)}.
\label{h2}
\end{equation}
$\delta_s$ is known for arbitrary values of $s$, but we will only use it 
for integer $s\in I\!\!N$ (cf.\ formula (\ref{3.27})).

For the parallel fluctuation continuum, we have  
$s_{\parallel}\equiv 1$ and 
$\frac{d\delta_1(k)}{dk}=-\frac{2}{1+k^2}$ so that
\begin{equation}
h_{\parallel}(a)=-\frac{1}{\pi}\int_{-\infty}^{\infty}
\frac{\ln\left(1-e^{-a\sqrt{1+k^2}}\right)}{1+k^2}dk.
\label{h3}
\end{equation}
This integral is convergent for $a>0$. For high temperatures we have
$a=\beta_T\mu\rightarrow 0$. Using the approximation
\begin{equation}
\ln\left(1-e^{-a\sqrt{1+k^2}}\right)\approx \ln\left(a\sqrt{1+k^2}\right)
=\frac12\ln\left(a^2+(ak)^2\right),
\label{h4}
\end{equation}
the substitution $\zeta=ak$ yields
\begin{equation}
h_{\parallel}(a)\approx -\frac{a}{2\pi}\int_{-\infty}^{\infty}
\frac{\ln\left(a^2+\zeta^2\right)}{a^2+\zeta^2}d\zeta=-\ln(2a)
\label{h5}
\end{equation}
which means
\begin{equation}
h_{\parallel}(\beta_T\mu)\sim - \ln(\beta_T\mu)>0 \quad \mbox{for} \quad
\beta_T\mu\longrightarrow 0.
\label{h6}
\end{equation}

In the case of perpendicular fluctuations, the behaviour of
the phase shift is much more complicated. We have 
$\delta_{s_{\perp}}(k)=\delta_{\perp}(k;\kappa^2)$ because (cf.\ eq.\ 
(\ref{5.29}))
\begin{equation}
s_{\perp}=s_{\perp}(k;\kappa^2)=\sqrt{\frac{25}{4}+4\kappa^2(1+k^2)}-\frac12.
\label{h7}
\end{equation}

The Skyrme--less limit ($\kappa^2=0$)
reduces this to $s_{\perp}(k;\kappa^2=0)=2$ and
$\frac{d\delta_2(k)}{dk}=-\frac{2}{1+k^2}-\frac{4}{4+k^2}$, 
and one can use the same approximations as in the parallel case
which yield
\begin{eqnarray}
h_{\perp}(a;\kappa^2=0) & = & -\frac{1}{\pi}\int_{-\infty}^{\infty}
\ln\left(1-e^{-a\sqrt{1+k^2}}\right)\left[\frac{1}{1+k^2}+
\frac{2}{4+k^2}\right] dk \nonumber \\
& \stackrel{\zeta=ak}{\approx} & -\frac{a}{2\pi}\int_{-\infty}^{\infty}
\ln\left(a^2+\zeta^2\right)\left[\frac{1}{a^2+\zeta^2}+
\frac{2}{4a^2+\zeta^2}\right] d\zeta \nonumber \\
& = & - \left(\ln(2a) + \ln(3a)\right)
\label{h7a}
\end{eqnarray}
and thus
\begin{equation}
h_{\perp}(a;\kappa^2=0) \sim - 2 \ln(\beta_T\mu)>0 \quad \mbox{for} \quad
\beta_T\mu\longrightarrow 0.
\label{h7b}
\end{equation}

Including the Skyrme term ($\kappa^2>0$) changes this behaviour. First,
in order to restrict the following 
discussion to positive $k$, we perform an
integration by parts in (\ref{h1}) and use the symmetry of the integrand
to obtain
\begin{equation}
h_{\perp}(a;\kappa^2)=\frac{1}{\pi}
\int_0^{\infty}\left[-\delta_{\perp}(k;\kappa^2)\right]\frac{d}{dk}
\left[\ln\left(1-e^{-a\sqrt{1+k^2}}\right)\right]dk.
\label{h8}
\end{equation}
For fixed $k$, $s_{\perp}(k;\kappa^2)$ and thus $-\delta_{\perp}(k;\kappa^2)$
increase with $\kappa^2$, so $h_{\perp}(a;\kappa^2)$ 
increases with increasing Skyrme coupling
$\kappa^2$ for fixed $a$.
Now the phase shift (\ref{3.26}), i.e.\  
\begin{equation}
\delta_{\perp}(k;\kappa^2)=\arg
\frac{\Gamma(-ik)\Gamma(1-ik)}{\Gamma\left(-ik+\frac12-\sqrt{\frac{25}{4}
+4\kappa^2(1+k^2)}\right)\Gamma\left(-ik+\frac12+\sqrt{\frac{25}{4}
+4\kappa^2(1+k^2)}\right)}
\label{h9}
\end{equation}
can be evaluated numerically. The main result is that for fixed $\kappa^2$, 
one can always find positive real numbers $\alpha=\alpha(\kappa^2)>1$, 
$c=c(\kappa^2)>0$ such that
\begin{itemize}
\item for $k\in(0,\alpha)$, $|\delta_{\perp}(k;\kappa^2)|$ is bounded by
$c<\infty$
\item for $k\ge\alpha$, $-\delta_{\perp}(k;\kappa^2)$ increases linearly.
\end{itemize}
It is this linear asymptotic behaviour of
$\delta_{\perp}(k;\kappa^2)=\CO(k)$ for $k\rightarrow\infty$ which 
determines the leading $a$--term in the asymptotic $a\rightarrow 0$--expansion
of (\ref{h8}). Here, we see one major influence of the Skyrme term on the
original $O(3)$--$\sigma$ model, because without Skyrme--term, we have
$\delta_{\perp}(k;\kappa^2=0)=\CO\left(\frac{1}{k^2}\right)$ 
for $k\rightarrow\infty$.

Besides a numerical analysis, one can also visualize the linear asymptotic 
behaviour of $\delta_{\perp}(k;\kappa^2)$ by some simple analytic arguments.
Consider the monotonically increasing series $(k_i)$ of real positive numbers 
for which $s_{\perp}(k_i;\kappa^2)\in I\!\!N$. For large $k_i$, we may estimate
$s_{\perp}(k_i;\kappa^2)\sim [\kappa k_i]$ where the Gauss bracket $[]$ denotes
the integer part of a real number. With
\begin{equation}
\arctan\left(\frac{1}{\kappa+1}\right) \le \arctan\left(\frac{k_i}{n}\right)
\le \frac{\pi}{2}, \qquad n\in\{1,\ldots,[\kappa k_i]\},
\label{h10}
\end{equation}
one can use (\ref{3.27}) to obtain
\begin{equation}
\arctan\left(\frac{1}{\kappa+1}\right)\cdot[\kappa k_i] \le
-\delta_{\perp}(k_i;\kappa^2) \le \frac{\pi}{2}\cdot([\kappa k_i]+1).
\label{h11}
\end{equation}
Thus, the series $(-\delta_{\perp}(k_i;\kappa^2))_{i\in I\!\!N}$ increases
linearly for $i\rightarrow\infty$. It should be possible to use the general
formula for $\delta_{\perp}(k_i;\kappa^2)$ to show that this is also true
for the interpolating values of $k$.

We may therefore use the approximation
\begin{equation}
-\delta_{\perp}(k;\kappa^2) \stackrel{\textstyle<}{\sim}
-\tilde{\delta}_{\perp}(k;\kappa^2) = \left\{
\begin{array}{ll}c(\kappa^2) & \mbox{for}\; 0<k<\alpha \\
m(\kappa^2)k+b(\kappa^2) & \mbox{for}\; k\ge\alpha
\end{array}\right.
\label{h12}
\end{equation}
with $m(\kappa^2)=\CO(\kappa)$ for $\kappa^2\rightarrow\infty$. The small
$\kappa^2$--behaviour of $m(\kappa^2)$ is difficult to estimate, but it is
clear from the above discussion that $m(\kappa^2)\rightarrow 0$,
$b(\kappa^2)\rightarrow 2\pi$ for $\kappa^2\rightarrow 0$; the Skyrme--less
limit thus changes the asymptotic $k$ behaviour of the phase shift
$\delta_{\perp}(k;\kappa^2)$ from a linear decrease to a constant value.

With (\ref{h12}), we estimate the integral (\ref{h8}):
\begin{eqnarray}
h_{\perp}(a;\kappa^2) & \stackrel{\textstyle<}{\sim} & \frac{1}{\pi}
\int_0^{\alpha}\left[-\tilde{\delta}_{\perp}(k;\kappa^2)\right]\frac{d}{dk}
\left[\ln\left(1-e^{-a\sqrt{1+k^2}}\right)\right]dk \nonumber \\
& & \qquad {} +\frac{1}{\pi}
\int_{\alpha}^{\infty}\left[-\tilde{\delta}_{\perp}(k;\kappa^2)\right]
\frac{d}{dk}
\left[\ln\left(1-e^{a\sqrt{1+k^2}}\right)\right]dk.
\label{h13}
\end{eqnarray}
The first integral is easy:
\begin{eqnarray}
\frac{c(\kappa^2)}{\pi}
\int_0^{\alpha}\frac{d}{dk}
\left[\ln\left(1-e^{-a\sqrt{1+k^2}}\right)\right]dk 
& = & \frac{c(\kappa^2)}{\pi}
\ln\left(\frac{1-e^{-a\sqrt{1+\alpha^2}}}{1-e^{-a}}\right) \nonumber \\
& = & \ln\left(\sqrt{1+\alpha^2}\right)+\CO(a) \quad \mbox{for}\; 
a\rightarrow 0.
\label{h14}
\end{eqnarray}
The second integral splits up into two parts,
\begin{equation}
\frac{m(\kappa^2)}{\pi}\int_{\alpha}^{\infty}k\frac{d}{dk}
\left[\ln\left(1-e^{-a\sqrt{1+k^2}}\right)\right]dk
+ \frac{b(\kappa^2)}{\pi}\int_{\alpha}^{\infty}\frac{d}{dk}
\left[\ln\left(1-e^{-a\sqrt{1+k^2}}\right)\right]dk
\label{h15}
\end{equation}
with
\begin{equation}
\int_{\alpha}^{\infty}\frac{d}{dk}
\left[\ln\left(1-e^{-a\sqrt{1+k^2}}\right)\right]dk = -
\ln\left(1-e^{-a\sqrt(1+\alpha^2)}\right) = 
\CO(\ln(a)) \quad \mbox{for}\; 
a\rightarrow 0.
\label{h16}
\end{equation}
In order to evaluate the remaining contribution, we insert
\begin{equation}
0<\frac{d}{dk}\ln\left(1-e^{-a\sqrt{1+k^2}}\right)=\frac{k}{\sqrt{1+k^2}}
\frac{a}{e^{a\sqrt{1+k^2}}-1}\stackrel{\textstyle<}{\sim}
\frac{a}{e^{ak}-1} \quad \mbox{for}\; k\gg 1
\label{h17}
\end{equation}
into the first integral of (\ref{h15}).
Using the dilogarithmic function
\begin{equation}
\mbox{dilog}(x):=\int_1^x\frac{\ln(t)}{1-t}dt
\label{h18}
\end{equation}
we obtain
\begin{eqnarray}
\int_{\alpha}^{\infty}k\frac{d}{dk}
\left[\ln\left(1-e^{-a\sqrt{1+k^2}}\right)\right]dk 
& \stackrel{\textstyle<}{\sim} & \int_{\alpha}^{\infty} 
\frac{ak}{e^{ak}-1}dk \nonumber \\
& = & \frac{1}{a}\int_{\alpha a}^{\infty}\frac{\zeta d\zeta}{e^{\zeta}-1}
\nonumber \\
& = & \frac{\pi^2}{6a}+\frac{1}{a}\mbox{dilog}\left(e^{\alpha a}\right)
+\frac12\alpha^2 a \nonumber \\
& = & \frac{\pi^2}{6a} - \alpha + \CO(a) \quad \mbox{for}\; 
a\rightarrow 0.
\label{h19}
\end{eqnarray}
Obviously, this is the leading term in the asymptotic 
$a\rightarrow 0$ expansion of $h_{\perp}(a;\kappa^2)$. Thus
\begin{equation}
h_{\perp}(\beta_T\mu;\kappa^2)\sim \frac{\pi m(\kappa^2)}{6}\frac{1}{\beta_T\mu}
\quad \mbox{for}\; \beta_T\mu\rightarrow 0.
\label{h20}
\end{equation}

In the Skyrme--less limit, this term vanishes ($m(\kappa^2=0)=0$), and the leading
$\beta_T\mu\rightarrow 0$ behaviour of 
$h_{\perp}(\beta_T\mu;\kappa^2\rightarrow 0)$ 
is given by (\ref{h7b}).

\end{appendix}

\end{document}